\documentclass[fleqn,usenatbib]{mnras}

\usepackage{newtxtext,newtxmath}
\usepackage[T1]{fontenc}
\usepackage{ae,aecompl}
\usepackage{pdflscape}

\usepackage{graphicx}	% Including figure files
\usepackage{amsmath}	% Advanced maths commands
\usepackage{amssymb}	% Extra maths symbols

\title[VVVX cluster candidates]{New Galactic Star Clusters Discovered in the Disk Area of the VVVX Survey.}

\author[J. Borissova et al.]{
J. Borissova,$^{1,2}$\thanks{Based on observations gathered with VIRCAM at the ESO VISTA telescope, as part of observing programs 198.B-2004.}
V.D. Ivanov,$^{3}$
P.W. Lucas,$^{4}$
R. Kurtev,$^{1,2}$
J. Alonso-Garcia$^{5,2}$
\newauthor
S. Ram\'irez Alegr\'ia$^{5},$
D. Minniti,$^{6,2}$
D. Froebrich,$^{7}$
M. Hempel,$^{2,8}$
N. Medina,$^{2,1}$
\newauthor
A.-N. Chen\'e,$^{9}$
and
M.A. Kuhn$^{2,10}$
\\
$^{1}$Instituto de F\'isica y Astronom\'ia, Universidad de Valpara\'iso, Av. Gran Breta\~na 1111, Playa Ancha, Casilla 5030, Chile.\\
$^{2}$Millennium Institute of Astrophysics (MAS), Santiago, Chile.\\
$^{3}$European Southern Observatory, Karl Schwarzschildstr. 2, D-85748 Garching bei M\"unchen, Germany\\
$^{4}$Centre for Astrophysics, University of Hertfordshire, College Lane, Hatffeld, AL10 9AB, UK.\\
$^{5}$Centro de Astronom\'{i}a (CITEVA), Universidad de Antofagasta, Avenida Angamos 601, Antofagasta, Chile \\
$^{6}$Departamento de F\'isica, Facultad de Ciencias Exactas, Universidad And\'es Bello, Av. Fernandez Concha 700, Las Condes,
Santiago, Chile.\\
$^{7}$Centre for Astrophysics and Planetary Science, School of Physical Sciences, University of Kent, Canterbury, CT2 7NH, UK\\
$^{8}${Instituto de Astrof\'isica, Pontificia Universidad Cat\'olica de Chile, Casilla 306, Santiago 22, Chile}\\
$^{9}$Gemini Observatory, Northern Operations Center, 670 A'ohoku Place, Hilo, HI 96720, USA\\
$^{10}$Department of Astronomy, California Institute of Technology, Pasadena, CA 91125, USA
}
\date{Accepted XXX. Received YYY; in original form ZZZ}

\pubyear{2018}

\begin{document}
\label{firstpage}
\pagerange{\pageref{firstpage}--\pageref{lastpage}}
\maketitle

\begin{abstract}
The ``VISTA Variables in the V\'{\i}a L\'actea eXtended (VVVX)'' ESO Public Survey is a near-infrared photometric sky survey that covers nearly 1700\,deg$^2$ towards the Galactic disk and bulge. It is well-suited to search for new open clusters, hidden behind dust and gas. The pipeline processed and calibrated $K_S$-band tile images of 40\% of the disk area covered by VVVX was visually inspected for stellar over-densities. Then, we identified cluster candidates by examination of the composite $JHK_S$ color images. The color-magnitude diagrams of the cluster candidates are constructed. Whenever possible the Gaia DR2 parameters are used to calculate the mean proper motions, radial velocities, reddening and distances. We report the discovery of 120 new infrared clusters and stellar groups. Approximately, half of them (47\%) are faint, compact, highly reddened, and they seem to be associated with other indicators of recent star formation, such as nearby Young Stellar Objects, Masers, \ion{H}{ii} regions or bubbles. 
The preliminary distance determinations allow us to trace the clusters up to 4.5 kpc, but most of the cluster candidates are centered at 2.2 kpc. The mean proper motions of the clusters, show that in general, they follow the disk motion of the Galaxy. 
\end{abstract}


%%%%%%%%%%%%%%%%%%%%%%%%%%%%%%%%%%%%%%%%%%%%%%%%%%

%%%%%%%%%%%%%%%%% BODY OF PAPER %%%%%%%%%%%%%%%%%%

\section{Introduction}

The Milky Way environment provides a unique place to test the predictions of cosmological models and the theories of galaxy formation. Moreover, our location within our own Galaxy gives us a close-up look at star clusters which in turn has implications on extragalactic star clusters studies with the next generation facilities like the NASA's James Webb Space Telescope (JWST) and the ESO's European Extremely Large Telescope (E-ELT). In preparation for these we have to complete the census of star clusters in the Galaxy, as well as to create template samples of well understood benchmark star clusters. This has motivated the renewed interest in star clusters, made possible by the new all-sky infrared (IR) surveys, such as 2MASS \citep[Two Micron All Sky Survey;][]{Skr06}, GLIMPSE \citep[Galactic Legacy Infrared Mid-Plane Survey Extraordinaire;][]{Ben03}, WISE \citep[Wide-field Infrared Survey Explorer;][]{Wri10}, VVV \citep[VISTA Variables in the V\'{\i}a L\'actea;][]{Min10} and UKIDSS GPS \citep[The Galactic Plane Survey;][]{Luc08}. These new IR surveys added many new objects to the traditional optical catalogs (e.g. WEBDA \cite{Dia02}; MWSC or Milky Way global survey of Star Clusters database - \citeauthor{Kha13} \citeyear{Kha13}, \citeauthor{Sch14} \citeyear{Sch14}), including some heavily obscured star clusters visible only in the infrared \citep[hereafter, IR clusters; see for example][]{Mor13, Zas13, Cam16}. Taking into account the resent work of \cite{Ryu18},  the total number of known infrared clusters has know exceeded 6300 objects.
 
Many of these new IR clusters are based on the ESO Large Public Survey VISTA Variables in the V\'ia L\'actea -- VVV\footnote{P.I. D. Minniti, \url{https://vvvsurvey.org}}, because it delivered deep sub-arcsec seeing $ZYJHK_S$ images and photometry \citep{Min10,Sai12}. The VVV survey mapped the IR variability of the Milky Way bulge and southern mid-plane over a period of six years (2010-2016). The VVV footprint contained about 300 known clusters, but more than 750 new clusters and candidate clusters were added in VVV-based works \citep[e.g.,][]{Bor11,Bor14,Sol14,Bar15,Iva17,Fro17}. Follow-up spectroscopy of some new candidates confirmed their cluster nature, improving the census of young massive clusters in the Galaxy: seven new ones were added and they contain at least one newly discovered WR star \citep{Che13,Che15, Her16}. We have reported a new, massive WR star \citep[>100 solar mass,][]{Che15}, a new, massive cluster in the far edge of the Galactic bar \citep{Ram14}, etc. The multi-epoch $K_S$ VVV observations also made it possible to investigate the IR variability of the cluster members \citep{Bor14,Bor16,Nav16, Med18}. 

The existence of these previously undiscovered open clusters was predicted by \cite{Por10}, but the VVV survey has also improved the census of globular clusters in the Galactic bulge and disk. The number of globular clusters in our Galaxy was estimated to be not more than 160 \citep[see][for an empirical approach to the missing globular clusters]{Iva05}. But recent discoveries in the VVV area reveal about 100 new globular cluster candidates \citep{Min11,Mon11,Bor14,Min17a,Min17b,Min17c}. If confirmed, this will drastically change the landscape of the old cluster population in our Galaxy.

A new project, ``VISTA Variables in the V\'{\i}a L\'actea eXtended'' (VVVX) survey, was launched in 2016 as an extension of the complied VVV survey in order to enhance its legacy value. The VVVX will spend a total of $\sim$2000\,hr of VISTA time over 3\,years to extend the VVV time-baseline and to nearly double the VVV spatial coverage up to $\sim$1700\,deg$^2$ from $l$=230$^\circ$ to $l$=20$^\circ$ (7$^{\rm h}$\,$<$\,$\alpha$\,$<$\,19$^{\rm h}$). Based on the VVV experience, our predictions are that VVVX will also significantly increase the number of star clusters in our Galaxy. Many are expected near the tangent point of the Carina arm region, which harbors very massive young clusters such as Westerlund\,2, NGC\,3603 and the Carina Nebula Complex. The extension of VVVX towards the third Galactic quadrant is fundamental to unveil new clusters along the Perseus arm and to trace the proposed Outer arm, reaching the edge of the Galactic disk. 

Thus, following \cite{Bor11} we expanded our investigation of the Milky Way cluster population to include the new areas covered by VVVX, aiming to improve the star cluster census and to continue building a statistically significant sample of clusters, with homogeneously derived parameters. We concentrate on objects that are practically invisible in the optical bands. The improved catalog of star clusters in the Galaxy can help to constrain theoretical models of cluster formation \citep[e.g.,][]{Pfa16} and to determine some fundamental relations between basic cluster parameters.

Two and half years after the start of the VVVX survey we have at our disposal only a part of the VVVX footprint. In this paper, we report the first results of our visual search for new star cluster candidates in the VVVX disk area. More specifically by April, 2018, appox. 55\%  of the new VVVX area was observed.  From this, we have searched 74\%, excluding only the extension of the VVVX bulge, which we plan to examine by automated tools. With respect to the whole new VVVX area the total area searched in this paper is 40\%.

%All papers should start with an Introduction section, which sets the work
%in context, cites relevant earlier studies in the field by \citet{Others2013},
%and describes the problem the authors aim to solve \citep[e.g.][]{Author2012}.

\section{Observations and Data Reduction}

The VVV and VVVX data were obtained with the 4.1-meter ESO VISTA telescope \cite[Visual and Infrared Survey Telescope for Astronomy;][]{Eme06} located at Cerro Paranal, Chile, with the 16-detector VIRCAM \citep[VISTA Infrared CAMera;][]{Dal06}. It has a $\sim$1$\,\times$\,1.5\,deg$^2$ field of view and works in the 0.9-2.5\,$\mu$m wavelength range and a pixel scale of 0.34\,arcsec\,px$^{-1}$. The data are reduced with the VISTA Data Flow System \cite[VDFS;][]{Irw04,Eme04} at the Cambridge Astronomical Survey Unit\footnote{\url{http://casu.ast.cam.ac.uk/}} (CASU). Processed images and photometric catalogs are available from the ESO Science Archive\footnote{\url{http://archive.eso.org/}} and from the VISTA Science Archive\footnote{\url{http://horus.roe.ac.uk/vsa/}} \citep[VSA;][]{Cro12}. A single VIRCAM image, called paw, contains large gaps; six paws taken in a spatial offset pattern must be combined to fill them in, obtaining a contiguous image, called tile. The VVVX paws and tiles are aligned along $l$ and $b$. The total exposure time of the tiles are: 8\,sec in $K_S$ (for a single epoch), 24\,sec in $H$ and 60\,in $J$ band. 
The tiles overlap by a few arc minutes in Galactic latitude and longitude -- just like with the VVV -- to ensure homogeneous coverage, spatial continuity and overall photometric and astrometric consistency. The tile overlaps result in a small fraction of duplicate sources, which is advantageous for variable stars, but that must be taken into account when analyzing maps or star counts that span wider areas than that of a single tile. Importantly, these tile overlaps allow to test the intrinsic accuracy of the photometry and astrometry. Therefore, the VVVX followed the same observing strategy as the VVV.
 
\begin{figure}
\begin{center}
   % \resizebox{\hsize}{!}\includegraphics{bulge-12102017.jpg}
\includegraphics[width=8cm]{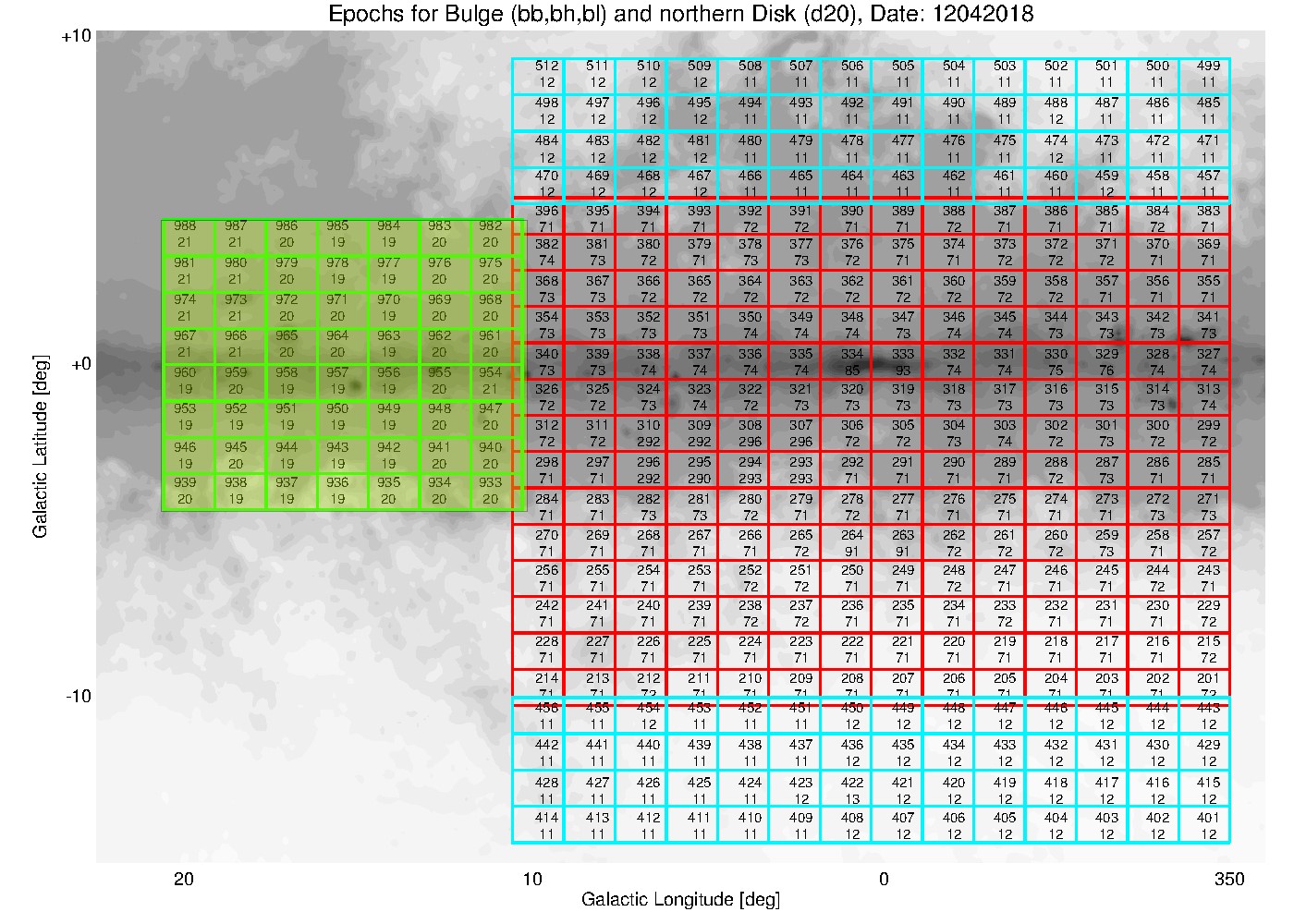}
\includegraphics[width=8cm]{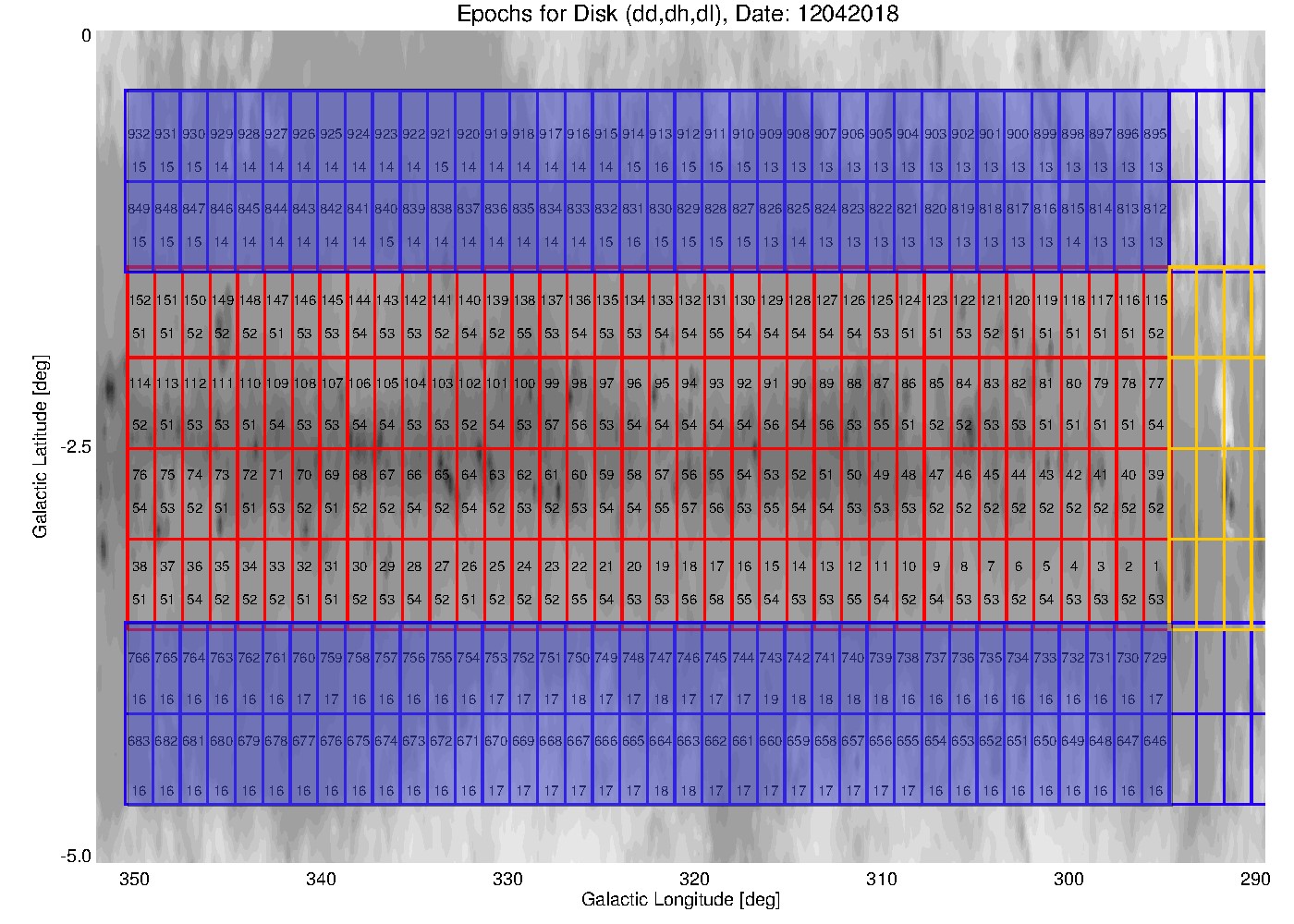}
\includegraphics[width=8cm]{MN180342f2.jpg}   
\end{center}
\caption{The VVVX Survey area. The red squares show the VVV Survey tiles, the green -- the Northern disk extension, the cyan -- the bulge area extension, the blue -- the extensions of both sides of the South VVV disk, and the yellow -- new VVVX outer disk fields. The searched areas are filled with green and dark blue colors. The numbers in each tile are its identification number (top) and the number of $K_S$ epochs by April 12, 2018.}
\label{tiles_id}
\end{figure}

Figure\,\ref{tiles_id} shows the VVV and VVVX footprints. In the bulge this corresponds to 20$^\circ$\,$\times$\,24$^\circ$ (14\,$\times$\,22 tiles). These are shown in Fig.\,\ref{tiles_id} in red for the former VVV tiles (b201 to b396), in cyan for the new southern bulge extension from tiles b401 to b456, and for the new northern bulge extension from tiles b457 to b512. The new northern disk area covers a 10$^\circ$\,$\times$\,9$^\circ$ patch (7\,$\times$\,8 tiles). These are shown in green in Fig.\,\ref{tiles_id}, from tiles e933 to e988. The new southern disk area covers 120$^\circ$\,$\times$\,9$^\circ$, split into two stripes of 83\,$\times$\,2 tiles each (Fig.\,\ref{tiles_id}; tiles e601 to e766, and e767 to e932 are marked in blue) and an extension along the Galactic mid-plane region by 65$^\circ$\,$\times$\,4$^\circ$, (Fig.\,\ref{tiles_id}, tiles e1001 to e1180 are shown in yellow). The original VVV-disk area is also shown in red in Fig.\,\ref{tiles_id}, consisting of tiles d001 to d152. During its first two and half years (2016 to 2018), the VVVX survey covered in the $JHK_S$ bands the southern bulge extension from tiles b401 to b456; e601 to e766, and e767 to e932. The northern disk tiles are from e933 to e988. 

\section{Cluster Search and Validation}

\begin{figure}
\begin{center}
\includegraphics[width=7.4cm]{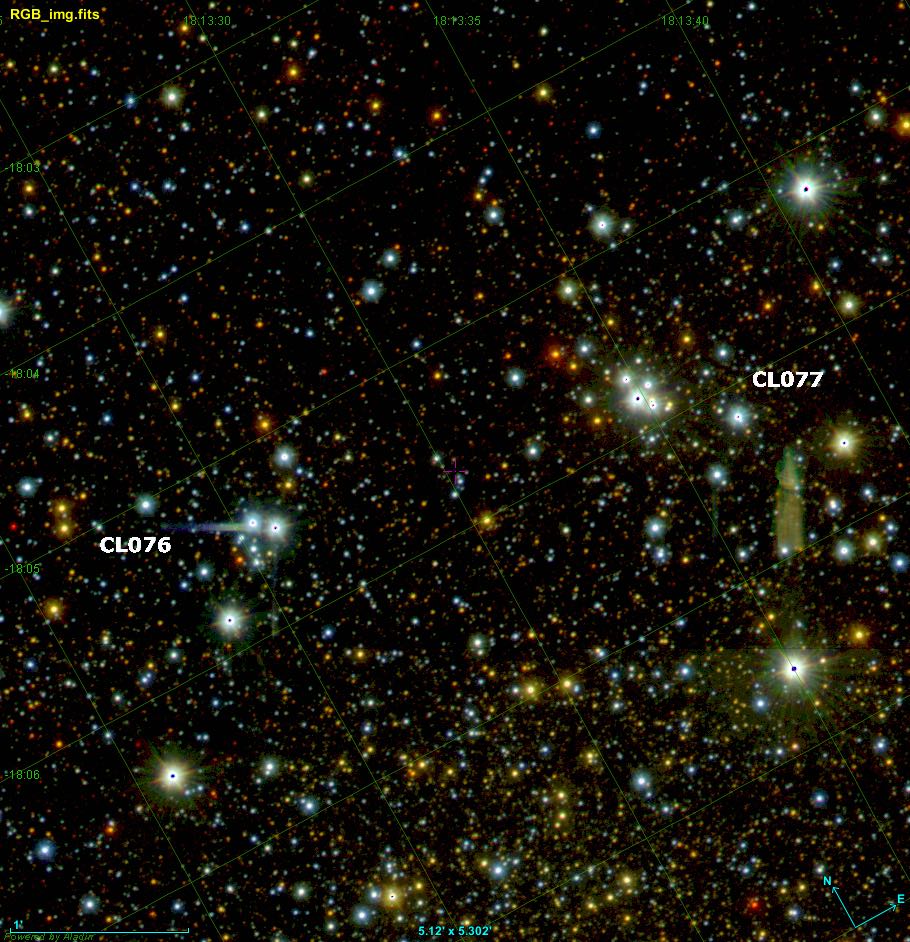}
\includegraphics[width=7.4cm]{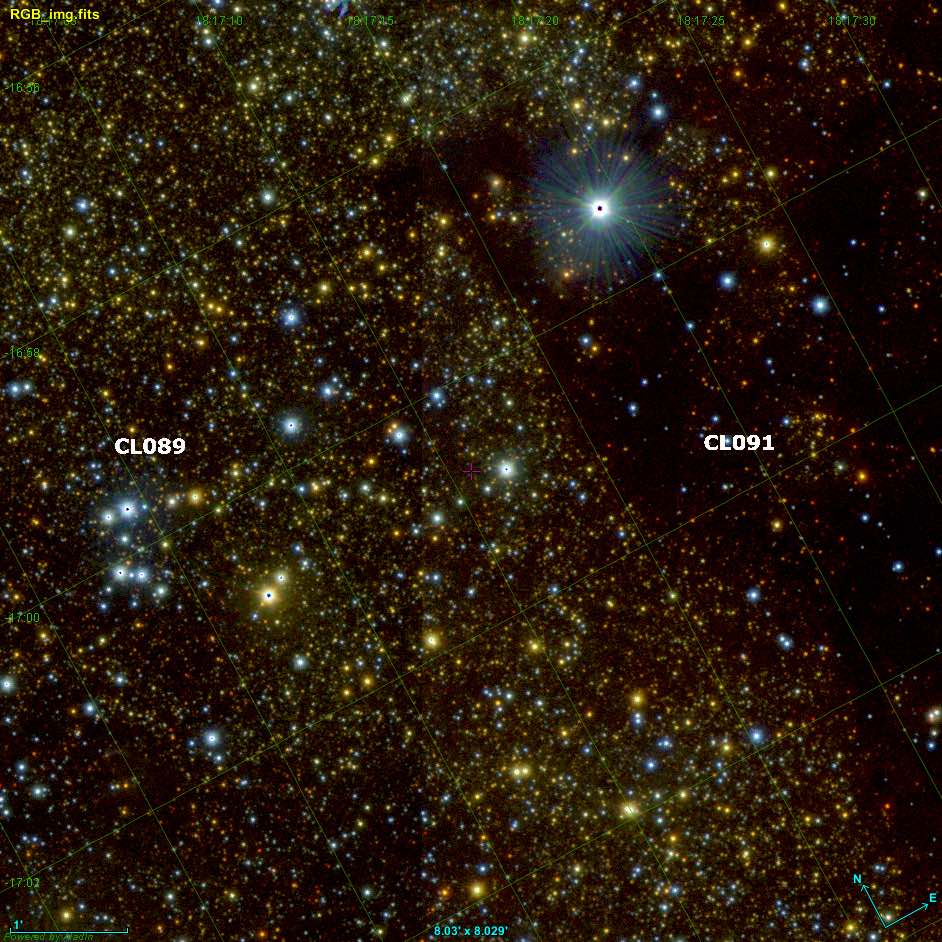}
\includegraphics[width=7.4cm]{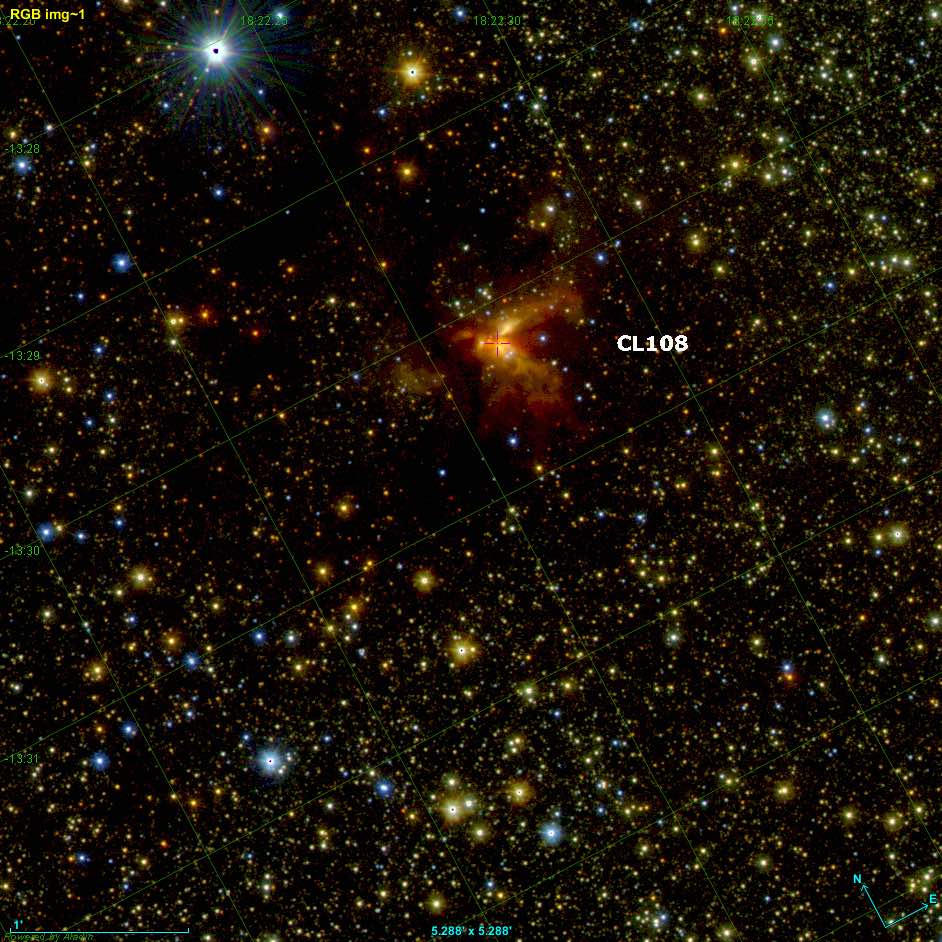}
\end{center}
\caption{VVVX $JHK_S$ composite color images of some open cluster candidates. The field of view is 5\,$\times$\,5\,arcmin.}
\label{rgb_typical}
\end{figure}

Our previous experience has shown that the number densities used by automated cluster searches \citep{Iva02,Bor03} tend to yield a large number of spurious cluster candidates in the inner Milky Way, because of the uneven distribution of the obscuring dust. They require time-consuming manual follow up at different, e.g. mid-IR wavelengths, and at the end they are not much more effective than simple visual searches. Furthermore, here we are aiming at relatively faint and heavily reddened clusters, undiscovered in the 2MASS, DENIS and GLIMPSE surveys. Therefore, we adopted the visual inspection as our search method. All observed images were retrieved from the CASU database and initially the $K_S$ tile images were visually inspected. The preliminary list of candidates was created on the basis of the detected local overdensity of the number of stars with respect to the surrounding area.    
Then, the composite $JHK_S$ color images were created and we verified the compact appearance, distinctive from the surrounding field and containing at least 5-6 stars with similar colors concentrated towards the objects' center. Figure\,\ref{rgb_typical} shows  the composite $JHK_S$ color images of some newly discovered cluster candidates and stellar groups for illustration. The color images of the whole sample are given in the Appendix~C. 

\begin{figure}
\includegraphics[width=\columnwidth]{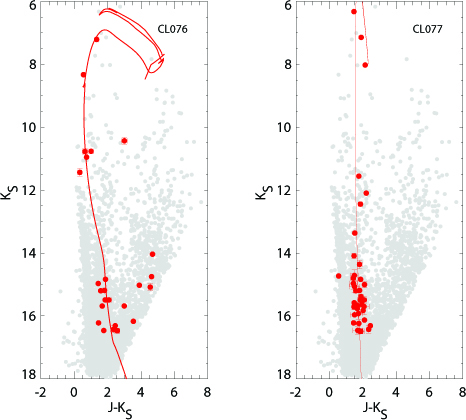}
\caption{The VVVX CL076 and CL077 $K_{S}$ vs $(J-K_{S})$ diagram, with most probable cluster members (red large circles) overplotted (see text). The solid red lines are best fit solar isochrones of 32 and 10 Myr, respectively, taken for Padova database. }
\label{cmd_examples}
\end{figure}

The next step in our validation process was the analysis of the color-magnitude diagrams. To construct them, we performed PSF photometry of a $2.5\,\times\,2.5$ arcmin area in the $J$, $H$, and $K_S$ bands surrounding the selected candidate. We used the Dophot  photometric routine following \cite{Alo17}. The instrumental magnitudes were transformed to the standard system, and the saturated stars (usually $K_S \leq 11.5$ mag, depending on crowding) were replaced by 2\,MASS stars \citep[2MASS PSC;][]{Skr06}. This procedure is described in detail by \cite{Alo17} and ~\cite{Bor11,Bor14}.  
Figure\,\ref{cmd_examples} shows two examples of color-magnitude diagrams for VVVX CL076 and VVVX CL077, both classified as open clusters. The most probable cluster members are selected by statistical decontamination procedure (for more details see \citealt{Bor11}) and Gaia DR2 proper motion diagrams (see next paragraph).  Following \cite{Zas13}, we then fit the Padova theoretical isochrones with solar metallicity \citep{Bre12} (http://stev.oapd.inaf.it/cgi-bin/cmd).  The best fit for VVVX CL076 is 32 Myr, while VVVX CL077 is most probably 10 Myr old. We used the Gaia RD2 parallaxes as initial values for the distances (see Table\,\ref{parallaxes}), while the mean reddening is determined in the interactive process of the fitting as $E(J-K_{S})=1.6\pm0.2$ and $E(J-K_{S})=1.7\pm0.2$, respectively. 
The color-magnitude diagrams of the whole sample, excluding very faint and compact candidates, are given in Appendix~D. 
The ages of the cluster candidates however will be a subject of follow-up paper, together with up-coming spectroscopic data. 
 
\begin{figure}
\includegraphics[width=6.7cm]{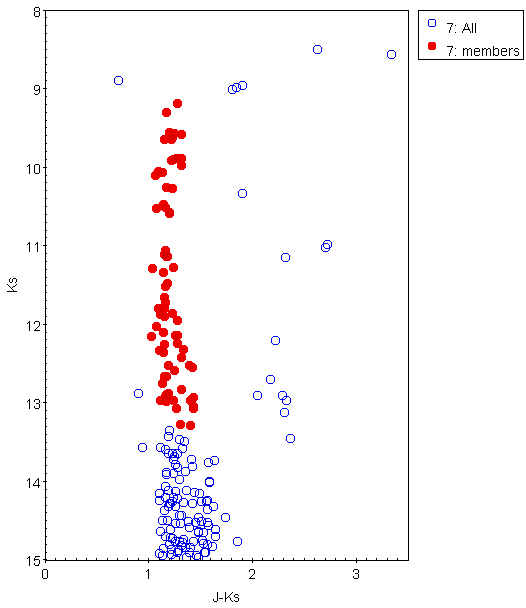}
\includegraphics[width=6.7cm]{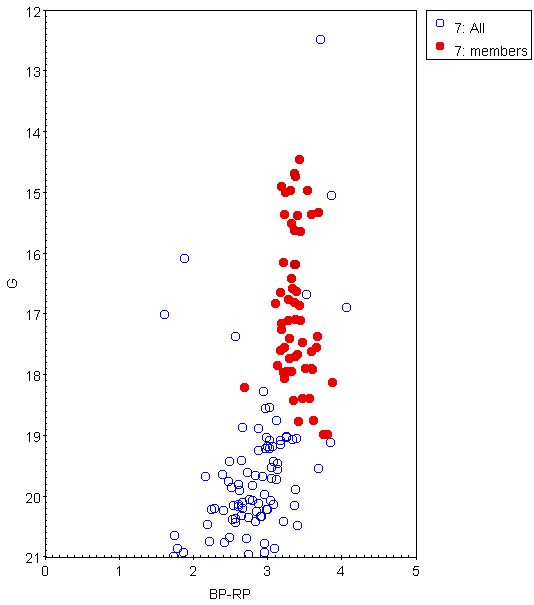}
\includegraphics[width=6.7cm]{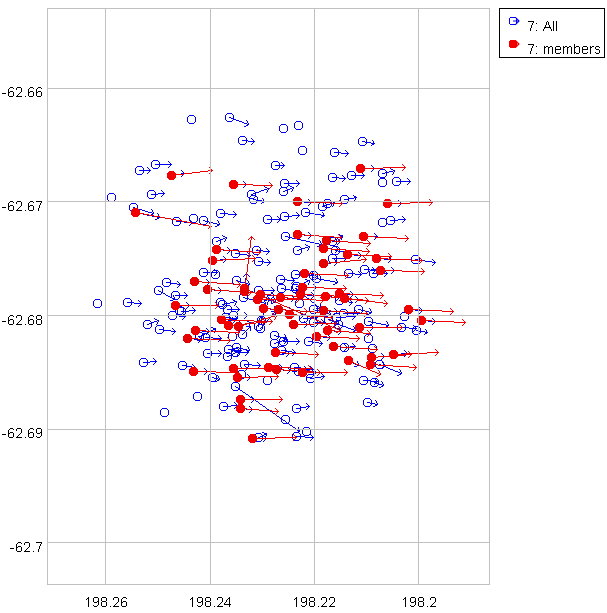}
\caption{The VVV and Gaia DR2 color magnitude and proper motion diagrams with sky vectors overplotted for known open cluster Danks\,2. Blue open circles are all stars in the selected area, the red solid ones stand for most probable cluster members.  The proper motion vectors are scaled by factor of 2 for visibility.} 
\label{pm}
\end{figure}

The Gaia astrometric mission was launched in December
2013 \citep{2016A&A...595A...1G} to measure positions, parallaxes, proper motions and photometry for over $10^9$ sources as well as to obtain physical parameters and radial velocities for millions of stars.  
Its recent Data Release 2 (Gaia DR2), has covered the initial
22 months of data taking \citep{gaia-collaboration2018}. As pointed out by \cite{2018arXiv180409381G}, it is expected that the members of the clusters span small range of distances and their members follow a common space motion which is, in general, different from the bulk of the field stars in the same region. This is illustrated in Fig.\,\ref{pm} (lower panel) for the massive open cluster Danks\,2. Thus, we can select the stars with common proper motion and their mean parallax can give us the distance to the cluster. In our case, only small number of probable cluster members have Gaia DR2 counterparts, because we are covering the infrared bands and fainter stars. Moreover, the cluster members are moving with the disk and it is hard to distinguish between both motions. 

Therefore, we adopted the following procedure: the most probable cluster members from the statistically decontaminated $K_{S}$ vs $(J-K_{S})$ diagrams (Fig.\,\ref{pm}, upper, red solid circles) are cross-matched with Gaia DR2 catalog (middle, red solid circles).
The proper motion vector diagrams $\mu_{\delta}$ vs $\mu_{\alpha}$$\cos\delta$  are created and the stars with obviously different proper motion are rejected. Whenever possible, the histograms of radial velocities are also examined and the outliers are rejected.
For such ``cleaned'' sample of probable cluster members we calculated the median value and the standard deviation of  $\mu_{\alpha}$$\cos\delta$ and $\mu_{\delta}$ (Col: 4 to 7 of Table\,\ref{parallaxes}), which we consider as the proper motion of the cluster.
Then, only the parallaxes with errors less than 20\%  are selected and used to calculate the median parallax and its standard deviation (Col: 8 and 9 of Table\,\ref{parallaxes}). No correction of  parallax zero-point is performed \citep{2018arXiv180409366L}.
Using the TOPCAT implementation of \cite{2015PASP..127..994B}; \cite{2016ApJ...832..137A} and \cite{2018arXiv180409376L} method we calculated ``the best estimate of distance using the Exponentially Decreasing Space Density prior'' and ``the 5th and 95th percentile confidence intervals''. The ``best estimate'' value is consider as the distance to the cluster, while the standard deviation of the 5th and 95th values is taken as the error of the determination. 

It is hard to compare the obtained distances, because there is no data in the literature for our objects. Indirectly, we can compare some kinematic distance estimates when the cluster candidate is projected close to an \ion{H}{ii} region. For example, the calculated Gaia distance of 2385 pc for VVVX CL038 is in reasonable agreement with the kinematic distance of 2024 pc of the associated \ion{H}{ii} region WRAY16-205 \citep{2008ApJ...689..194S}.
Nevertheless, we caution that such calculated distances should be taken as preliminary.  
 
We have also collected $K$-band spectra of 3 and 2 stars for VVVX CL010 and CL011, respectively, using the IR spectrograph and imaging camera SofI in long-slit mode, mounted on the ESO New Technology Telescope (NTT) \footnote{{Based on observations gathered with ESO program 0101.C-0519(A).}}. The instrument set-up give a resolution of R=2200, and the total exposure times were 900 and 1200\,s. The reduction procedure for the spectra is described in \cite{2012A&A...545A..54C}. The equivalent widths (EWs) were measured from the continuum-normalized spectra using the {\sc iraf}\footnote{IRAF is distributed by the National Optical Astronomy Observatory, which is operated by the Association of Universities for Research in Astronomy (AURA) under cooperative agreement with the National Science Foundation.} task {\it splot}. 

\begin{figure}
\includegraphics[width=\columnwidth]{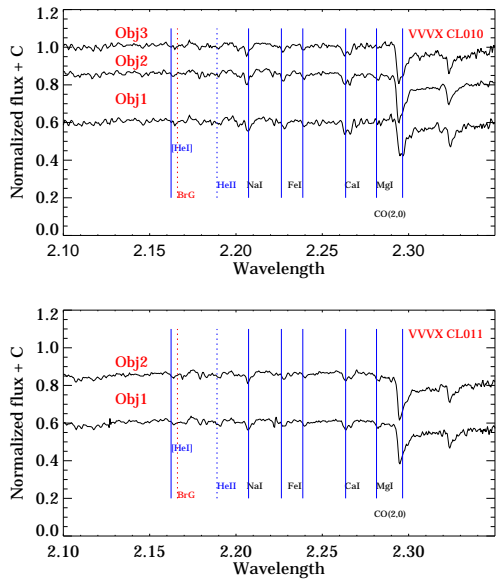}
\vspace{-2cm}
\caption{SofI low resolution spectra of VVVX CL010 and CL011.}
\label{spectra}
\end{figure}

Fig.\,\ref{spectra} shows the continuum-normalized spectra, all clearly exhibited characteristic features of cool stars, such as \ion{Na}{i} (2.21 $\mu$m); \ion{Ca}{i}  (2.26 $\mu$m) atomic line blends, and the first band-head of CO (2.29 $\mu$m).  Following  \cite{Bor14} and using the EWs of these three lines and \citep{2001AJ....122.1896F} calibration we calculated the mean metallicity of ${\rm [Fe/H]}=-0.39\pm0.06$ for CL010 and ${\rm [Fe/H]}=-0.02\pm0.11$ for CL011.

Some parameters of these stars are measured in the Gaia DR2 catalog and are tabulated in  Table\,\ref{specrtal_par}. To test $T_{\rm eff}$ as a proxy of distance, we obtained the spectral types for every star by comparing the Gaia temperature with the Straizys et al. (1981) Sp. Type vs Temperature calibration, and then the individual extinctions and distance moduli are calculated by spectroscopic parallaxes using the intrinsic colors and luminosities, again from Straizys et al. (1981). Thus, the distance moduli to CL010 and Cl011, calculated as a mean value of the individual measurements are $(M-{\rm m})_0=12.61\pm1.07$ (3.43 kpc) and $(M-{\rm m})_0=12.01\pm0.61$ (2.56 kpc), respectively. 
A comparison with the Gaia distances given in Table\,\ref{parallaxes} shows good agreement. Therefore, in some cases, when we have poorly measured Gaia parallaxes, but individual Gaia temperatures, it is still possible to estimate the distances using spectroscopic parallaxes.

\begin{landscape}
\begin{table}
\centering
\caption{Parameters of stars with spectra}
\begin{tabular}{llllrrllcrllcc} % 13 columns, alignment for each
		\hline
\multicolumn{1}{c}{Name}	&	\multicolumn{1}{c}{$\alpha(2000)$}	              &	  \multicolumn{1}{c}{$\delta(2000)$}  &	\multicolumn{1}{c}{$\pi$}	 	&	\multicolumn{1}{c}{$\mu_\alpha\cos\delta$} & \multicolumn{1}{c}{$\mu_\delta$} &	\multicolumn{1}{c}{Gmag}                  &	\multicolumn{1}{c}{$T_{\rm eff}$}	&	\multicolumn{1}{c}{Sp type}	&	 \multicolumn{1}{c}{$J$}	&	\multicolumn{1}{c}{$H$}	 	&	\multicolumn{1}{c}{$K_S$} & \multicolumn{1}{c}{$E(J-K)$}	&          \multicolumn{1}{c}{Distance}	\\

\multicolumn{1}{c}{}	&	\multicolumn{1}{c}{$^\circ$}	              &	  \multicolumn{1}{c}{$^\circ$}  &	\multicolumn{1}{c}{mas}	 	&	\multicolumn{1}{c}{mas} & \multicolumn{1}{c}{mas} &	\multicolumn{1}{c}{mag}                  &	\multicolumn{1}{c}{K}	&	\multicolumn{1}{c}{}	&	 \multicolumn{1}{c}{mag}	&	\multicolumn{1}{c}{mag}	 	&	\multicolumn{1}{c}{mag} & \multicolumn{1}{c}{mag}	&          \multicolumn{1}{c}{kpc}	\\
\hline 
\multicolumn{14}{c}{CL010}\\
Obj1	&	192.66456	&	-64.91643	&	0.0092$\pm$0.056	&	-6.202$\pm$0.076	&	0.144$\pm$0.070	&	13.47$\pm$0.001	&3849.86	&	M0	&9.79$\pm$0.03	&	8.70$\pm$0.05	&	8.24$\pm$0.03	&	 0.92	&          2.31	\\
Obj2	&	192.66222	&	-64.91845	&	0.1428$\pm$0.109	&	-5.832$\pm$0.151	&	0.233$\pm$0.136	&	12.93$\pm$0.003	&3296.00	&	M5	&8.73$\pm$0.03	&	7.60$\pm$0.02	&	7.10$\pm$0.02	&	 0.99	&          3.99	\\
Obj3	&	192.65980	&	-64.92115	&	0.0624$\pm$0.056	&	-6.280$\pm$0.076	&	-0.707$\pm$0.070	&	14.38$\pm$0.001	&3823.60	&	M1	&10.69$\pm$0.03	&	9.56$\pm$0.04	&	9.14$\pm$0.02	&	 0.88	&          3.98	\\
\multicolumn{14}{c}{CL011}\\
Obj1	&	192.81099	&	-60.59321	&	0.4624$\pm$0.037	&	-14.310$\pm$0.045&	-1.730$\pm$0.048	&	11.80$\pm$0.000	& 4048.41&          K6	&9.61$\pm$0.03	&	8.89$\pm$0.03	&	8.70$\pm$0.02	&	 0.31	&          3.00	\\
Obj2	&	192.80710	&	-60.59457	&	0.3658$\pm$0.062	&	-6.740$\pm$0.080	&	-0.260$\pm$0.082	&	11.52$\pm$0.002	& 3669.00&	M3	&8.29$\pm$0.02	&	7.28$\pm$0.04	&	6.94$\pm$0.02	&	 0.68	&          2.14         \\
\hline
\end{tabular}
\label{specrtal_par}
\end{table}
\end{landscape}

\section{The Catalog} 

We identified 120 new candidate star clusters or stellar groups, listed in Table\,\ref{candidates}. The first column gives the identification, followed by the equatorial coordinates of the center determined by eye, eye-ball measured apparent cluster radius in arcsec, the name of the corresponding VVVX tile and some comments about the nature of the object such as: presence or absence of nebulosity or \ion{H}{ii} region around the cluster, known nearby infrared, radio and X-ray sources, young stellar objects (YSO), outflow candidates and masers, taken from the SIMBAD database (\url{http://simbad.u-strasbg.fr/simbad/}). 

During the visual inspection we recovered 69 known star clusters, mainly from the lists of \cite{Bic03}, \cite{Mer05} and  \citet{Sol12}, as well as some clusters from WEBDA\footnote{\url{http://www.univie.ac.at/webda/}} \citep{Dia02} database. These are not all known clusters in the VVVX area, many of optically visible  clusters (for examples those from \citet{Kha13}) are not taken in consideration. Another thirteen, previously unknown  candidates are in common with \cite{Luc18} catalog. All these objects are removed from the catalog presented here. 
Finally, \cite{Ryu18} published a WISE catalog of 923 new star cluster candidates.    Comparison shows that we have 7 common candidates. The VVVX cluster candidates CL065; CL070; CL073; CL074; CL084; CL097 and CL120 are matched within 20 arcsec radius with Ryu670; Ryu657; Ryu687; Ryu664; Ryu715; Ryu711; Ryu796, respectively. Since the \cite{Ryu18} catalog was published after this paper was submitted to the journal, we consider these objects simultaneously discovered and didn't remove them from our list, as in the case of previous catalogs. 
 
The preliminary analysis, based only on the appearance on the VVVX images shows  two general groups: 53\% (64 objects) have the typical appearance of open clusters, the rest of the sample (56 objects) are projected on H\,II regions, bubbles, nebulosity or around some early OB stars, YSOs and IR sources. 

\begin{figure}
\includegraphics[width=\columnwidth]{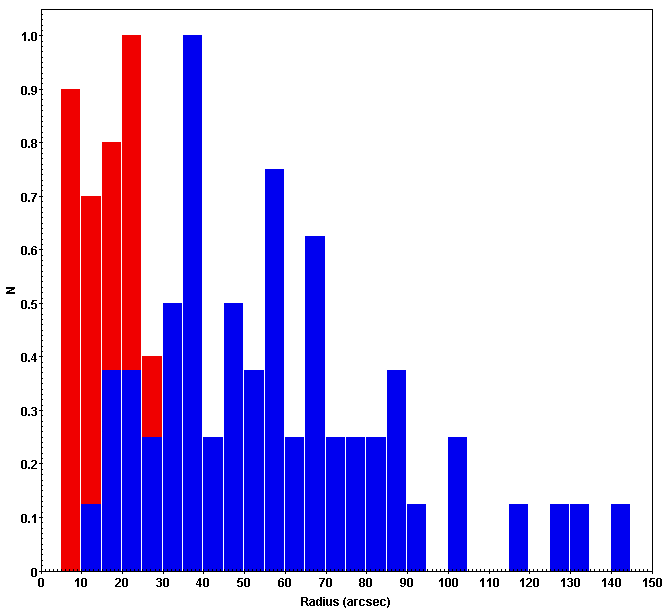}
\includegraphics[width=\columnwidth]{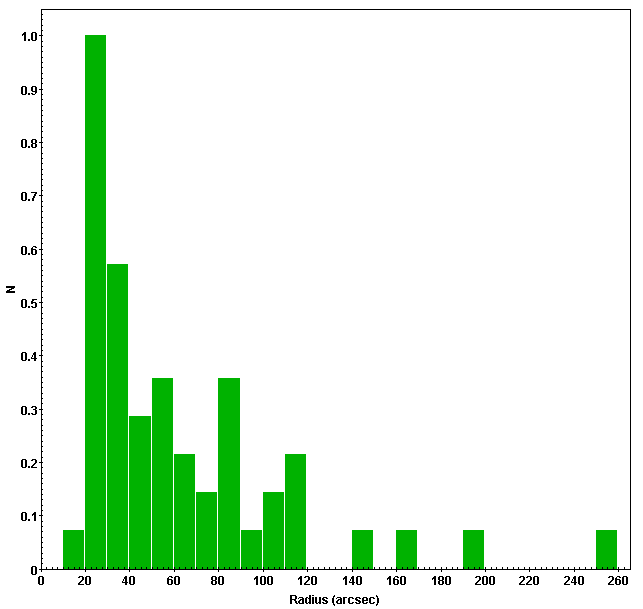} 
\caption{Distribution of the detected objects with measured cluster radius (given in arcsec). The red color stands for young/embedded clusters, while the blue one is for open ones. The second panel shows know clusters in the VVVX area.}
\label{his_radius}
\end{figure}

The cluster radii were measured by eye on the VVVX $K_S$ tiles. This method was preferred over automated algorithms, in order to include objects that are not resolved into stars. The area around cluster candidates is smoothed and the density contours are over-plotted with the lower limit of the contour equal to the density of comparison field.  The normalized histogram of the Number of star clusters vs Radius is shown in Fig.\,\ref{his_radius}. The sample was divided by two: embedded/young clusters (red color) and open cluster candidates (blue color). The Gaussian distribution gives the mean radius of the young clusters sample of $18.5\pm9$\,arcsec, while the open cluster sample is centered on $57\pm30$\,arcsec. 
For comparison, the Gaussian distribution of the known infrared clusters in the investigated VVVX area (see previous paragraph) is centered on $62\pm47$\,arcsec. Thus, in this study, we are adding the fainter and compact (with small angular sizes) new candidates. This looks very similar to the size distribution plot for the clusters found by visual inspection of the GPS/VVV area \citep{Fro17}. The small sizes suggest that the clusters are distant, but the sample could be biased and further investigation is necessary to confirm or reject such suggestion. 
    
\begin{figure*}
\includegraphics[width=8.5cm]{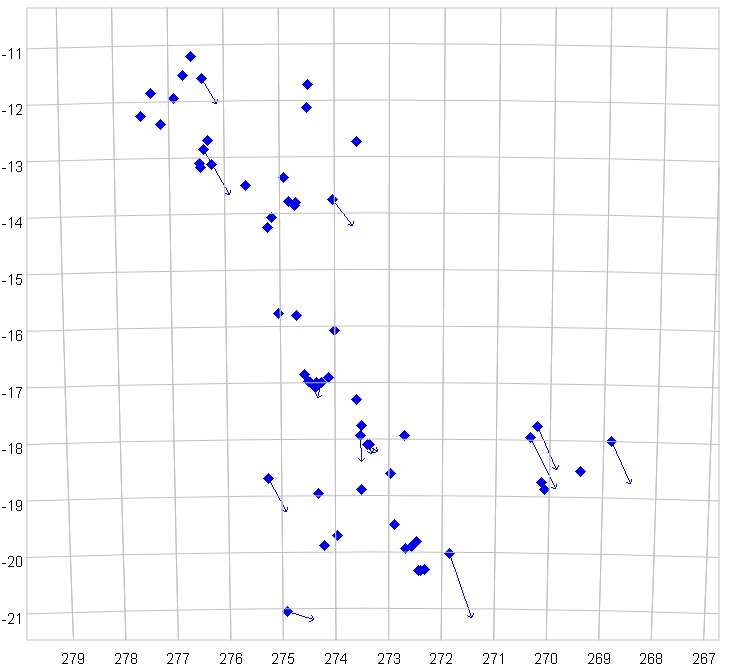}
\includegraphics[width=8.6cm]{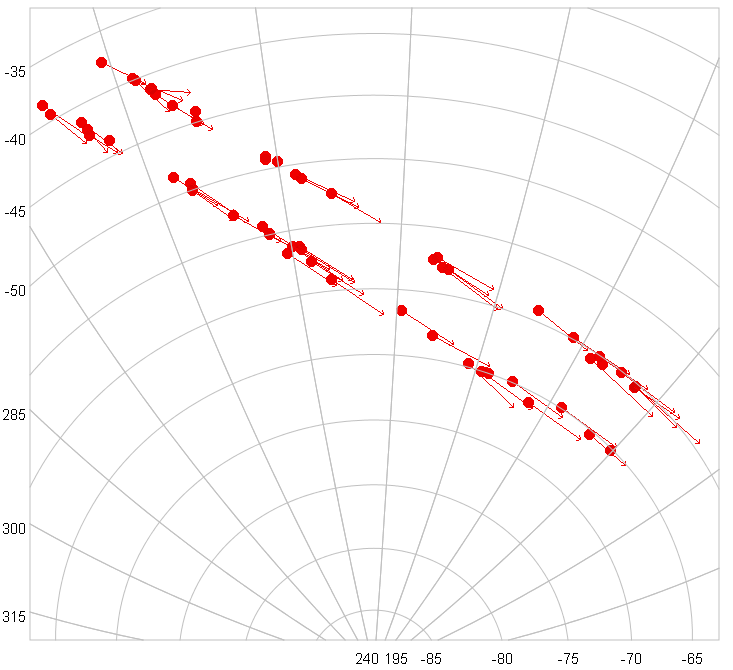}
\includegraphics[width=9cm]{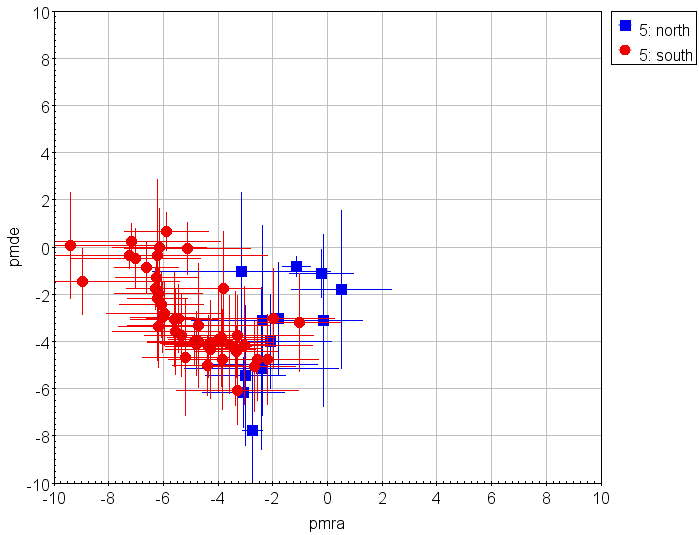}
\caption[]{ The diagram of mean proper motions of the cluster candidates. The blue squares are for the candidates projected in North disk VVVX area, while red circles stand for the Southern part. The proper motion vectors are scaled by factor of 2 for visibility. }.
\label{pm_cl}
\end{figure*}

Using the information from Table\,\ref{specrtal_par} (see previous paragraph) we can derive some preliminary relations from the new sample of cluster candidates. 
In Fig.\,\ref{pm_cl}, left and middle part, we overplot the sky proper motion vectors, while the right panel shows the proper motion diagram. The blue squares and red  circles  represent the cluster candidates projected on the North and South VVVX disk area. As can be seen from the figure, in general the clusters follow the disk motion (with exception of CL049, 056, 080, 089 and 103). This is more notable in the South disk area, at higher Galactic latitudes, where the distribution seems more homogeneous. The last panel shows a clear concentration of the candidates towards the lower left part, where the galactic disk stars are projected.

\begin{figure*}
\includegraphics[width=\textwidth]{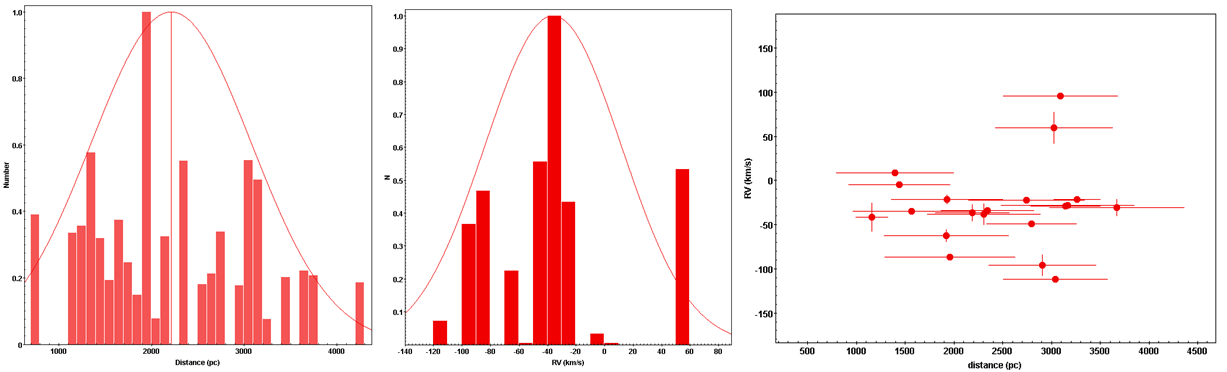}
\caption[]{ The histograms of distance and radial velocity distribution. The right panel shows RV vs Distance relation. The solid lines are Gaussian distribution.}
\label{dis_rv}
\end{figure*}

Figure \,\ref{dis_rv} illustrates the mean normalized distances and radial velocities of the cluster candidates, taking in consideration the corresponding errors. The distance distribution seems complex, with several groups around 1.3; 2.3 and 3.1 kpc. The mean value of the sample is $2.21\pm0.86$ kpc, calculated as a peak of the Gaussian distribution. Thus, Gaia DR2 distances can be used up to 3.5-4 kpc, but  most of the clusters are closer than 2 kpc. The radial velocities are distributed around $\rm R_V=-36\pm47$\,km/s.  A very week correlation (9\%) between radial velocities and distances is found, but this could be selection, small sample effect. 
Again, these values should be taken as preliminary and used with caution.

\section{Summary}

In this work we report a catalog of 120 new infrared clusters and stellar groups projected in the disk area covered by the ongoing ``VISTA Variables in the V\'{\i}a L\'actea eXtended (VVVX)'' ESO Public Survey. The search is performed by visual inspection on the pipeline processed and calibrated $K_S$-band tile images.  The initial list of candidates is then validated using the composite $JHK_S$ color images, $K_{S}$ vs $(J-K_{S})$ color-magnitude diagrams and Gaia DR2 proper motions. A smaller proportion of embedded candidates is detected in comparison with previous VVV searches, as would be expected given that much of the area searched is slightly further from the Galactic equator. The smaller number density of candidates is explained by the fact that the area covered by this new search (the outer part of the Galactic disc) is significantly less crowded and reddened than that of the previous search. In general, the clusters follow the disk motion. The Gaia DR2 distances are estimated up to 3.5-4 kpc, but  most of the clusters are closer than 2 kpc.

\section*{Acknowledgements}

We gratefully acknowledge data from the ESO Public Survey program ID 198.B-2004 taken with the VISTA telescope, and products from the Cambridge Astronomical Survey Unit (CASU). This work has made use of data from the European Space Agency (ESA) mission {\it Gaia} (\url{https://www.cosmos.esa.int/gaia}), processed by the {\it Gaia} Data Processing and Analysis Consortium (DPAC, \url{https://www.cosmos.esa.int/web/gaia/dpac/consortium}). Funding for the DPAC has been provided by national institutions, in particular the institutions participating in the {\it Gaia} Multilateral Agreement. Support is provided by the Ministry for the Economy, Development and Tourism, Programa Iniciativa Cientica Milenio grant IC120009, awarded to the Millennium Institute of Astrophysics (MAS). JB and RK thank ESO for the financial support during their Jan 2018 visit. ANC's work is supported by the Gemini Observatory, which is operated by the Association of Universities for Research in Astronomy, Inc., on behalf of the international Gemini partnership of Argentina, Brazil, Canada, Chile, and the United States of America.  SRA thanks the support by the FONDECYT Iniciaci\'on project No. 11171025 and the CONICYT + PAI ``Concurso Nacional Inserci\'on de Capital Humano Avanzado en la Academia 2017'' project PAI 79170089. J.A-G. also acknowledges support by FONDECYT Iniciaci\'on 11150916. MH acknowledges support by the BASAL Center for Astrophysics and Associated Technologies (CATA) through grant PFB-06. We thank anonymous referee for useful comments and suggestions.

%%%%%%%%%%%%%%%%%%%%%%%%%%%%%%%%%%%%%%%%%%%%%%%%%%



%\bibliographystyle{mnras}

\newpage
\bibliographystyle{mnras}

%%%%%%%%%%%%%%%%%%%%%%%%%%%%%%%%%%%%%%%%%%%%%%%%%%

%%%%%%%%%%%%%%%%% APPENDICES %%%%%%%%%%%%%%%%%%%%%
\appendix

\section{VVVX Cluster Candidates: Catalog}

In Table\,\ref{candidates} are tabulated 120 new candidate star clusters or stellar groups. The first column gives the identification, followed by the equatorial coordinates of the center determined by eye, eye-ball measured apparent cluster radius in arcsec, the name of the corresponding VVVX tile and some comments about the nature of the object.

\begin{landscape}
\begin{table}
\caption{VVVX Cluster Candidates: Catalog}
\begin{tabular}{llcccl} % four columns, alignment for each
		\hline
\multicolumn{1}{c}{Name}	&	\multicolumn{1}{c}{$\alpha(2000)$}	              &	  \multicolumn{1}{c}{$\delta(2000)$}	&	\multicolumn{1}{c}{Radius}	&	\multicolumn{1}{c}{VVVX tile}	&	\multicolumn{1}{c}{Comments}	\\
\multicolumn{1}{c}{}	&	\multicolumn{1}{c}{$^\circ$}	              &	  \multicolumn{1}{c}{$^\circ$} & arcsec & & \\
		\hline
VVVX CL001	&	180.380770	&	-65.137202	&	18	&	e0730	&	very faint, embedded group of stars					\\
VVVX CL002	&	180.397080	&	-65.126113	&	20	&	e0730	&	very faint, embedded group of stars					\\
VVVX CL003	&	184.276940	&	-60.220145	&	29	&	e0814	&	small open cluster					\\
VVVX CL004	&	184.307420	&	-60.303649	&	55	&	e0814	&	open cluster					\\
VVVX CL005	&	184.324520	&	-60.242285	&	45	&	e0814	&	small reddened open cluster					\\
VVVX CL006	&	185.437370	&	-65.356478	&	55	&	e0732	&	O9 star+group of bright stars					\\
VVVX CL007	&	187.473880	&	-60.114671	&	120	&	e0815	&	open cluster					\\
VVVX CL008	&	190.768300	&	-60.437844	&	34	&	e0816	&	faint, embedded group of stars					\\
VVVX CL009	&	191.765130	&	-60.072564	&	66	&	e0816	&	several bright stars, open cluster, IRAS 12441-5947					\\
VVVX CL010	&	192.658350	&	-64.916684	&	38	&	e0734	&	concentrated open cluster?					\\
VVVX CL011	&	192.800920	&	-60.592911	&	55	&	e0817	&	several bright stars, open cluster					\\
VVVX CL012	&	196.764630	&	-59.841214	&	62	&	e0818	&	several bright stars, open cluster					\\
VVVX CL013	&	198.576150	&	-65.840076	&	50	&	e0735	&	several bright stars, open cluster					\\
VVVX CL014	&	202.949120	&	-64.935656	&	55	&	e0737	&	open cluster					\\
VVVX CL015	&	203.442400	&	-59.141331	&	31	&	e0821	&	several bright stars, open cluster					\\
VVVX CL016	&	207.530690	&	-65.007603	&	12	&	e0738	&	very faint, weak nebula, group of stars					\\
VVVX CL017	&	208.820520	&	-65.056893	&	66	&	e0738	&	open cluster					\\
VVVX CL018	&	211.590200	&	-64.731412	&	23	&	e0739	&	small group, dissolved?					\\
VVVX CL019	&	217.911420	&	-58.051811	&	145	&	e0826	&	open cluster					\\
VVVX CL020	&	218.723750	&	-63.222653	&	36	&	e0742	&	open cluster					\\
VVVX CL021	&	218.823650	&	-57.981169	&	40	&	e0826	&	several bright stars, open cluster					\\
VVVX CL022	&	219.642140	&	-57.257272	&	80	&	e0826	&	open cluster					\\
VVVX CL023	&	220.166310	&	-57.441524	&	105	&	e0827	&	open cluster					\\
VVVX CL024	&	224.245480	&	-61.592499	&	28	&	e0744	&	open cluster					\\
VVVX CL025	&	234.033390	&	-52.575296	&	59	&	e0833	&	open cluster					\\
VVVX CL026	&	234.949620	&	-59.136058	&	68	&	e0748	&	B8Ib+group					\\
VVVX CL027	&	237.487450	&	-57.608643	&	70	&	e0749	&	group of stars					\\
VVVX CL028	&	237.510440	&	-51.179911	&	85	&	e0835	&	open cluster					\\
VVVX CL029	&	238.130460	&	-50.768071	&	125	&	e0835	&	open cluster					\\
VVVX CL030	&	238.709700	&	-56.558402	&	67	&	e0750	&	open cluster					\\
VVVX CL031	&	238.944100	&	-56.279803	&	140	&	e0750	&	OB star+group of stars					\\
VVVX CL032	&	239.836430	&	-56.233403	&	51	&	e0750	&	B9 star + group of stars					\\
VVVX CL033	&	240.062400	&	-49.524395	&	13	&	e0836	&	faint, weak nebula, group of stars, 2MASX J16001524-4931255					\\
VVVX CL034	&	240.695290	&	-56.720018	&	37	&	e0750	&	open cluster					\\
VVVX CL035	&	241.357520	&	-48.991222	&	38	&	e0837	&	very red group of stars, radio source [CAB2011] G332.40+2.46					\\
VVVX CL036	&	241.423240	&	-49.191992	&	17	&	e0837	&	embedded group+ Maser: Caswell OH 332.295+02.280?+YSO: 2MASS J16054183-4911294					\\
VVVX CL037	&	242.630320	&	-54.937447	&	8	&	e0752	&	embedded faint group + YSO [MHL2007] G328.9716-02.4664 1					\\
VVVX CL038	&	242.670130	&	-54.961111	&	21	&	e0752	&	group in \ion{H}{ii} region: WRAY 16-205 					\\
VVVX CL039	&	243.306690	&	-54.287039	&	22	&	e0752	&	open cluster, concentrated					\\
VVVX CL040	&	246.636880	&	-52.874636	&	60	&	e0754	&	compact group of 2MASS variable stars					\\
VVVX CL041	&	247.898680	&	-44.174854	&	33	&	e0841	&	open cluster					\\
VVVX CL042	&	248.132730	&	-44.924463	&	64	&	e0841	&	embedded +Variable Star of FU Ori type: V* V346 Nor					\\
VVVX CL043	&	250.190380	&	-43.120268	&	89	&	e0843	&	open cluster					\\
VVVX CL044	&	250.562580	&	-49.889192	&	72	&	e0756	&	open cluster					\\
\hline
\end{tabular}
\label{candidates}
\end{table}
\end{landscape}

\begin{landscape}
\begin{table}
\contcaption{VVVX Cluster Candidates: Catalog}
\begin{tabular}{llcccl} % four columns, alignment for each
		\hline
\multicolumn{1}{c}{Name}	&	\multicolumn{1}{c}{$\alpha(2000)$}	              &	  \multicolumn{1}{c}{$\delta(2000)$}	&	\multicolumn{1}{c}{Radius}	&	\multicolumn{1}{c}{VVVX tile}	&	\multicolumn{1}{c}{Comments}	\\
\multicolumn{1}{c}{}	&	\multicolumn{1}{c}{$^\circ$}	              &	  \multicolumn{1}{c}{$^\circ$} & arcsec & & \\
		\hline
VVVX CL045	&	250.705140	&	-50.061132	&	75	&	e0756	&	open cluster					\\
VVVX CL046	&	250.706130	&	-49.467184	&	50	&	e0757	&	GC?					\\
VVVX CL047	&	251.500630	&	-41.766043	&	59	&	e0844	&	open cluster					\\
VVVX CL048	&	251.723100	&	-41.233356	&	10	&	e0844	&	small group in cloud: MSX6C G343.0500+02.6094					\\
VVVX CL049	&	251.860180	&	-41.244139	&	15	&	e0844	&	group around Emission-line Star: [OSP2002] BRC 82 11					\\
VVVX CL050	&	252.328640	&	-48.592074	&	17	&	e0758	&	concentrated group around Object of unknown nature: XZLJ Nor 111					\\
VVVX CL051	&	253.042590	&	-40.065037	&	81	&	e0845	&	open cluster					\\
VVVX CL052	&	253.294640	&	-39.836432	&	31	&	e0846	&	open cluster					\\
VVVX CL053	&	255.785070	&	-37.520748	&	43	&	e0848	&	open cluster					\\
VVVX CL054	&	257.723300	&	-43.705101	&	30	&	e0762	&	embedded group					\\
VVVX CL055	&	259.483430	&	-42.657861	&	38	&	e0763	&	open cluster					\\
VVVX CL056	&	259.499020	&	-42.094804	&	86	&	e0764	&	open cluster					\\
VVVX CL057	&	259.780310	&	-41.405093	&	12	&	e0764	&	concentrated group					\\
VVVX CL058	&	262.414960	&	-39.466213	&	94	&	e0766	&	open cluster					\\
VVVX CL059	&	262.844920	&	-38.469595	&	100	&	e0766	&	open cluster					\\
VVVX CL060	&	268.852770	&	-18.000075	&	55	&	e0982	&	open cluster					\\
VVVX CL061	&	269.430960	&	-18.542233	&	31	&	e0975	&	group of stars  in  Interstellar matter: [GMG2004] 85					\\
VVVX CL062	&	270.094130	&	-18.869869	&	18	&	e0975	&	embedded cluster? Too big to be single YSO: [MJR2015] 113					\\
VVVX CL063	&	270.144320	&	-18.754833	&	29	&	e0975	&	group in \ion{H}{ii} region: MSX6C G010.5067+02.2285					\\
VVVX CL064	&	270.249460	&	-17.749453	&	69	&	e0975	&	open cluster					\\
VVVX CL065	&	270.369110	&	-17.946281	&	65	&	e0975	&	open cluster					\\
VVVX CL066	&	271.856700	&	-20.028048	&	36	&	e0961	&	open cluster					\\
VVVX CL067	&	272.318750	&	-20.312052	&	47	&	e0954	&	young cluster part of W31?					\\
VVVX CL068	&	272.405310	&	-20.320490	&	21	&	e0954	&	small group in Bubble  [CWP2007] CN 145					\\
VVVX CL069	&	272.436630	&	-20.336289	&	53	&	e0954	&	group in \ion{H}{ii} region: [L89b] 10.190-00.426					\\
VVVX CL070	&	272.482890	&	-19.806053	&	24	&	e0954	&	group in Bubble: [SPK2012] MWP1G010670-002100S					\\
VVVX CL071	&	272.579460	&	-19.902798	&	22	&	e0954	&	small group in \ion{H}{ii} region:  [AAJ2015] G010.630-00.338					\\
VVVX CL072	&	272.678330	&	-19.942037	&	14	&	e0954	&	group in \ion{H}{ii} region: HRDS G010.638-00.434					\\
VVVX CL073	&	272.713320	&	-17.931154	&	28	&	e0962	&	group in \ion{H}{ii} region: GAL 012.42+00.50					\\
VVVX CL074	&	272.884650	&	-19.510605	&	28	&	e0954	&	small group around YSO: IRAS 18085-1931					\\
VVVX CL075	&	272.971860	&	-18.605660	&	10	&	e0955	&	small group in \ion{H}{ii} region:  [KC97c] G011.9+00.0					\\
VVVX CL076	&	273.354540	&	-18.092123	&	17	&	e0955	&	concentrated group					\\
VVVX CL077	&	273.396040	&	-18.096607	&	53	&	e0955	&	concentrated group close to CL076+WR, binary cluster?					\\
VVVX CL078	&	273.504900	&	-18.891444	&	22	&	e0955	&	young cluster in \ion{H}{ii} region GRS G011.94 -00.62					\\
VVVX CL079	&	273.510870	&	-17.758839	&	20	&	e0955	&	group in Bubble: [SPK2012] MWP1G012930-000811					\\
VVVX CL080	&	273.527110	&	-17.934716	&	19	&	e0955	&	small group around B0V star  [MCF2015] 13, neb, concentrated					\\
VVVX CL081	&	273.587430	&	-12.742477	&	20	&	e0973	&	small group around YSO: MSX6C G017.3765+02.2512					\\
VVVX CL082	&	273.601520	&	-17.293137	&	18	&	e0956	&	group in Bubble: [SPK2012] MWP1G013382+000659					\\
VVVX CL083	&	273.955800	&	-19.703903	&	23	&	e0947	&	small very red group, neb. 					\\
VVVX CL084	&	274.006270	&	-16.085923	&	15	&	e0964	&	neb, IR: IRAS 18131-1606					\\
VVVX CL085	&	274.006350	&	-16.085481	&	17	&	e0964	&	group of stars, neb., IR source					\\
VVVX CL086	&	274.040020	&	-13.755454	&	49	&	e0972	&	open cluster					\\
VVVX CL087	&	274.111360	&	-16.913282	&	35	&	e0956	&	open cluster					\\
VVVX CL088	&	274.208750	&	-19.889306	&	10	&	e0947	&	small group around YSO: MSX6C G011.3757-01.6770					\\
\hline
\end{tabular}
%\label{candidates}
\end{table}
\end{landscape}

\begin{landscape}
\begin{table}
\contcaption{VVVX Cluster Candidates: Catalog}
\begin{tabular}{llcccl} % four columns, alignment for each
		\hline
\multicolumn{1}{c}{Name}	&	\multicolumn{1}{c}{$\alpha(2000)$}	              &	  \multicolumn{1}{c}{$\delta(2000)$}	&	\multicolumn{1}{c}{Radius}	&	\multicolumn{1}{c}{VVVX tile}	&	\multicolumn{1}{c}{Comments}	\\
\multicolumn{1}{c}{}	&	\multicolumn{1}{c}{$^\circ$}	              &	  \multicolumn{1}{c}{$^\circ$} & arcsec & & \\
		\hline
VVVX CL089	&	274.247920	&	-16.999377	&	48	&	e0956	&	open cluster					\\
VVVX CL090	&	274.312720	&	-18.954494	&	132	&	e0948	&	open cluster					\\
VVVX CL091	&	274.342440	&	-16.999377	&	25	&	e0956	&	young cluster or dust window					\\
VVVX CL092	&	274.361490	&	-17.092749	&	10	&	e0956	&	embedded group around YSO:[PW2010] 48					\\
VVVX CL093	&	274.461220	&	-16.999192	&	10	&	e0956	&	embedded, compact group in \ion{H}{ii} region: MSX6C G014.0329-00.5155					\\
VVVX CL094	&	274.477580	&	-11.725361	&	13	&	e0973	&	second small group of stars close to BDS9, sub-cluster?					\\
VVVX CL095	&	274.481180	&	-16.975518	&	10	&	e0956	&	small group around YSO: SSTGLMA G014.0632-00.5199					\\
VVVX CL096	&	274.492270	&	-12.123610	&	16	&	e0973	&	small group around YSO: IRAS 18151-1208					\\
VVVX CL097	&	274.561750	&	-16.851552	&	34	&	e0956	&	neb, YSOs, bubbles, IR					\\
VVVX CL098	&	274.696620	&	-15.814823	&	20	&	e0957	&	group in \ion{H}{ii} region: IRAS 18159-1550					\\
VVVX CL099	&	274.709510	&	-13.815189	&	15	&	e0965	&	small group in \ion{H}{ii} region: MSX6C G016.9512+00.7806, close to M16, sub-cluster?					\\
VVVX CL100	&	274.717510	&	-13.859019	&	16	&	e0965	&	4 bright stars, very  compact, part of M16, RSG?					\\
VVVX CL101	&	274.827630	&	-13.794123	&	78	&	e0965	&	group in M16 (NGC6611)  region, sub-cluster?					\\
VVVX CL102	&	274.916667	&	-13.362500	&	11	&	e0966	&	group in \ion{H}{ii} region: MSX6C G017.4507+00.8118					\\
VVVX CL103	&	274.923010	&	-21.059604	&	32	&	e0940	&	group					\\
VVVX CL104	&	275.026400	&	-15.778958	&	10	&	e0957	&	small group in \ion{H}{ii} region: [SPE2008b] IRAS 18171-1548 VLA 1					\\
VVVX CL105	&	275.147920	&	-14.070781	&	13	&	e0965	&	embedded source in \ion{H}{ii} region: MSX6C G016.9261+00.2854					\\
VVVX CL106	&	275.230480	&	-14.258462	&	10	&	e0965	&	small group around YSO: 2MASS J18205529-1415306					\\
VVVX CL107	&	275.250670	&	-18.696785	&	48	&	e0948	&	open cluster					\\
VVVX CL108	&	275.611970	&	-13.504312	&	45	&	e0966	&	young cluster in SFR: RAFGL 2136 					\\
VVVX CL109	&	276.228570	&	-13.120082	&	35	&	e0950	&	GC?					\\
VVVX CL110	&	276.294290	&	-12.704624	&	18	&	e0959	&	small group in \ion{H}{ii} region: [ABB2014] WISE G018.657-00.057					\\
VVVX CL111	&	276.365960	&	-12.853166	&	77	&	e0959	&	open cluster					\\
VVVX CL112	&	276.389290	&	-11.588813	&	20	&	e0967	&	open cluster					\\
VVVX CL113	&	276.427310	&	-13.172107	&	52	&	e0950	&	group in \ion{H}{ii} region: [CKW87] 182253.2-131203					\\
VVVX CL114	&	276.451510	&	-13.108666	&	11	&	e0959	&	small group around YSO: IRAS 18229-1308					\\
VVVX CL115	&	276.594240	&	-11.202625	&	20	&	e0967	&	group around emission line stars: LS IV -11 19					\\
VVVX CL116	&	276.738040	&	-11.535798	&	10	&	e0967	&	compact group around YSO: IRAS 18241-1134					\\
VVVX CL117	&	276.910160	&	-11.945382	&	12	&	e0960	&	compact group in \ion{H}{ii} region: [WBH2005] G019.608-0.237					\\
VVVX CL118	&	277.149220	&	-12.407517	&	10	&	e0960	&	compact group in \ion{H}{ii} region: IRAS 18257-1226					\\
VVVX CL119	&	277.312510	&	-11.839633	&	10	&	e0960	&	group in \ion{H}{ii} region: MSX6C G019.8817-00.5347					\\
VVVX CL120	&	277.511290	&	-12.257561	&	20	&	e0960	&	group in bubble: [SPK2012] MWP1G019605-009038\\
\hline
\end{tabular}
%\label{candidates}
\end{table}
\end{landscape}

\section{VVVX Cluster Candidates: Parameters.}

In Table\,\ref{parallaxes} are tabulated the candidate star clusters with reliable Gaia DR2 parameters.   The first column gives the identification, followed by the equatorial coordinates of the center, median proper motion and parallax, as well as the calculated distance and radial velocity, with corresponding errors. 

\newpage
\begin{landscape}\tabcolsep=5pt
\begin{table}
\caption{VVVX Cluster Candidates: Parameters}
\begin{tabular}{llcccccccccccccc} % 16 columns, alignment for each
		\hline
\multicolumn{1}{c}{Name}	&	\multicolumn{1}{c}{$\alpha(2000)$}	&	\multicolumn{1}{c}{$\delta(2000)$}	& \multicolumn{1}{c}{$\mu_\alpha\cos\delta$}	&	\multicolumn{1}{c}{Err $\mu_\alpha\cos\delta$}	&	\multicolumn{1}{c}{$\mu_\delta$}	&	\multicolumn{1}{c}{Err $\mu_\delta$}	&	\multicolumn{1}{c}{$\pi$}	&	\multicolumn{1}{c}{Err $\pi$}	&	\multicolumn{1}{c}{Dis} 	&	\multicolumn{1}{c}{Dis 5\%}	&	\multicolumn{1}{c}{Dis 95\%}	&	\multicolumn{1}{c}{RV} 	&	\multicolumn{1}{c}{Err RV}	&	\multicolumn{1}{c}{E(BP-RP)}	&	\multicolumn{1}{c}{Err E(BP-RP)}	\\
\multicolumn{1}{c}{}	&	\multicolumn{1}{c}{$^\circ$}	&	\multicolumn{1}{c}{$^\circ$}	& \multicolumn{1}{c}{mas}	&	\multicolumn{1}{c}{mas}	&	\multicolumn{1}{c}{mas}	&	\multicolumn{1}{c}{mas}	&	\multicolumn{1}{c}{mas}	&	\multicolumn{1}{c}{mas}	&	\multicolumn{1}{c}{kpc} 	&	\multicolumn{1}{c}{kpc}	&	\multicolumn{1}{c}{kpc}	&	\multicolumn{1}{c}{km s$^{-1}$} 	&	\multicolumn{1}{c}{km s$^{-1}$}	&	\multicolumn{1}{c}{mag}	&	\multicolumn{1}{c}{mag}	\\
\hline
VVVX CL003	&	184.276940	&	-60.220145	&	-6.2048	&	4.0251	&	-0.3767	&	3.2011	&	0.4006	&	0.2471	&	1568	&	1142	&	2838	&	-35.24	&	3.26	&	0.63	&	0.40	\\
VVVX CL004	&	184.307420	&	-60.303649	&	-5.8743	&	1.5290	&	0.6543	&	0.8239	&	0.2025	&	0.0713	&	3166	&	2543	&	4465	&	-27.99	&	6.44	&	0.80	&	0.21	\\
VVVX CL005	&	184.324520	&	-60.242285	&	-9.3827	&	3.2809	&	0.0499	&	2.2410	&	0.4337	&	0.1749	&	1710	&	1299	&	2855	&		&		&	0.78	&	0.43	\\
VVVX CL006	&	185.437370	&	-65.356478	&	-5.1240	&	2.3100	&	-0.0600	&	1.1000	&	0.3605	&	0.0772	&	2346	&	1921	&	3266	&	-34.39	&	0.52	&	0.64	&	0.40	\\
VVVX CL007	&	187.473880	&	-60.114671	&	-7.1671	&	3.2500	&	0.2260	&	0.7475	&	0.1850	&	0.0530	&	3670	&	3010	&	4958	&	-30.71	&	19.12	&	0.82	&	0.26	\\
VVVX CL009	&	191.765130	&	-60.072564	&	-6.1444	&	1.7354	&	-0.0144	&	1.6379	&	0.5445	&	0.2224	&	1439	&	1077	&	2555	&	-4.85	&	2.20	&	0.98	&	0.63	\\
VVVX CL010	&	192.658350	&	-64.916684	&	-7.2231	&	1.2559	&	-0.3623	&	0.5297	&	0.2183	&	0.0534	&	3409	&	2811	&	4584	&		&		&	0.65	&	0.36	\\
VVVX CL011	&	192.800920	&	-60.592911	&	-8.9430	&	3.5489	&	-1.4580	&	1.3995	&	0.4141	&	0.0683	&	2186	&	1832	&	2907	&	-36.81	&	18.53	&	0.63	&	0.38	\\
VVVX CL012	&	196.764630	&	-59.841214	&	-7.0325	&	2.3901	&	-0.4942	&	1.2560	&	0.2470	&	0.0646	&	3024	&	2466	&	4174	&	59.38	&	34.84	&	1.01	&	0.21	\\
VVVX CL013	&	198.576150	&	-65.840076	&	-6.6136	&	1.1734	&	-0.8688	&	1.1200	&	0.1952	&	0.0303	&	4257	&	3656	&	5297	&		&		&	0.63	&	0.13	\\
VVVX CL014	&	202.949120	&	-64.935656	&	-6.2479	&	1.4922	&	-1.3088	&	0.8171	&	0.2526	&	0.1972	&	1933	&	1425	&	3282	&		&		&	0.79	&	0.18	\\
VVVX CL015	&	203.442400	&	-59.141331	&	-6.2920	&	1.0532	&	-1.7710	&	0.8923	&	0.2979	&	0.0193	&	3260	&	2980	&	3651	&	-21.63	&	3.85	&	0.75	&	0.20	\\
VVVX CL017	&	208.820520	&	-65.056893	&	-6.1841	&	1.5941	&	-1.9572	&	1.0552	&	0.2501	&	0.0589	&	3092	&	2544	&	4199	&	95.32	&	1.10	&	0.69	&	0.21	\\
VVVX CL018	&	211.590200	&	-64.731412	&	-5.6120	&	1.6243	&	-3.0585	&	1.9840	&	0.2964	&	0.0315	&	3143	&	2763	&	3768	&	-29.33	&	0.89	&	0.64	&	0.06	\\
VVVX CL019	&	217.911420	&	-58.051811	&	-6.0215	&	1.1670	&	-2.9814	&	1.4978	&	0.2657	&	0.0456	&	3195	&	2697	&	4131	&		&		&	1.25	&	0.39	\\
VVVX CL020	&	218.723750	&	-63.222653	&	-6.2019	&	1.0804	&	-2.1719	&	2.6371	&	0.9905	&	1.1153	&	795	&	480	&	2224	&		&		&	1.07	&	0.16	\\
VVVX CL021	&	218.823650	&	-57.981169	&	-6.1757	&	1.4781	&	-3.3479	&	1.7196	&	0.3039	&	0.0834	&	2527	&	2034	&	3621	&		&		&	1.35	&	0.11	\\
VVVX CL022	&	219.642140	&	-57.257272	&	-6.0517	&	1.5384	&	-2.4212	&	1.4175	&	0.8489	&	0.0980	&	1158	&	998	&	1458	&	-41.37	&	32.49	&	1.12	&	0.22	\\
VVVX CL023	&	220.166310	&	-57.441524	&	-5.9583	&	2.1346	&	-2.8311	&	1.3637	&	0.3710	&	0.1483	&	1925	&	1477	&	3096	&	-21.68	&	10.59	&	1.15	&	0.25	\\
VVVX CL024	&	224.245480	&	-61.592499	&	-5.4323	&	1.1612	&	-3.0226	&	1.4221	&	0.2279	&	0.0997	&	2660	&	2083	&	3962	&		&		&	0.87	&	0.21	\\
VVVX CL025	&	234.033390	&	-52.575296	&	-4.7031	&	2.5805	&	-3.3457	&	2.6372	&	0.1983	&	0.0446	&	3781	&	3151	&	4969	&		&		&	0.94	&	0.17	\\
VVVX CL026	&	234.949620	&	-59.136058	&	-4.8548	&	1.7112	&	-4.0185	&	0.9792	&	0.4375	&	0.3185	&	1394	&	989	&	2693	&	8.72	&	0.34	&	0.97	&	0.38	\\
VVVX CL027	&	237.487450	&	-57.608643	&	-4.7041	&	1.3876	&	-3.9946	&	1.3720	&	0.2847	&	0.1872	&	1923	&	1428	&	3240	&	-62.72	&	13.12	&	0.94	&	0.24	\\
VVVX CL028	&	237.510440	&	-51.179911	&	-5.3555	&	1.3411	&	-3.7328	&	1.7328	&		&		&	 	&		&		&	-64.80	&	1.58	&	1.11	&	0.38	\\
VVVX CL029	&	238.130460	&	-50.768071	&	-5.5530	&	2.2960	&	-3.5770	&	1.8099	&	0.3526	&	0.1424	&	1993	&	1532	&	3176	&		&		&	1.37	&	0.07	\\
VVVX CL030	&	238.709700	&	-56.558402	&	-4.7720	&	1.3078	&	-4.0830	&	1.3743	&	0.2294	&	0.1996	&	1960	&	1441	&	3328	&	-87.11	&	6.31	&	0.87	&	0.26	\\
VVVX CL031	&	238.944100	&	-56.279803	&	-4.7880	&	1.2216	&	-4.1505	&	1.1470	&	0.3110	&	0.0526	&	2798	&	2356	&	3655	&	-49.13	&	0.27	&	0.98	&	0.20	\\
VVVX CL032	&	239.836430	&	-56.233403	&	-4.3580	&	1.5724	&	-4.2170	&	1.3106	&	0.3185	&	0.1039	&	2309	&	1823	&	3451	&	-38.27	&	23.88	&	0.99	&	0.32	\\
VVVX CL034	&	240.695290	&	-56.720018	&	-4.2690	&	2.2917	&	-4.0300	&	1.6007	&	0.2762	&	0.0611	&	2903	&	2393	&	3944	&	-95.56	&	23.94	&	0.93	&	0.03	\\
VVVX CL038	&	242.670130	&	-54.961111	&	-5.1820	&	1.5626	&	-4.6810	&	2.4419	&	0.2286	&	0.1289	&	2385	&	1827	&	3720	&		&		&	0.72	&	0.08	\\
VVVX CL039	&	243.306690	&	-54.287039	&	-4.2715	&	1.6418	&	-4.3275	&	2.0885	&	0.4372	&	0.1931	&	1653	&	1242	&	2827	&		&		&		&		\\
VVVX CL040	&	246.636880	&	-52.874636	&	-3.8840	&	1.9633	&	-3.8520	&	2.0800	&		&		&	 	&		&		&	-48.65	&	3.54	&	1.12	&	0.29	\\
VVVX CL043	&	250.190380	&	-43.120268	&	-3.2785	&	2.7690	&	-3.7650	&	1.6912	&		&		&	 	&		&		&	-59.46	&	0.24	&	1.43	&	0.28	\\
VVVX CL044	&	250.562580	&	-49.889192	&	-4.3970	&	1.2967	&	-5.0395	&	1.2528	&		&		&	 	&		&		&		&		&	1.24	&	0.10	\\
VVVX CL045	&	250.705140	&	-50.061132	&	-3.0395	&	2.4861	&	-4.1620	&	2.5077	&	0.5739	&	0.1870	&	1460	&	1120	&	2462	&		&		&	1.46	&	0.01	\\
VVVX CL046	&	250.706130	&	-49.467184	&	-3.3370	&	1.5034	&	-4.4080	&	1.1117	&		&		&	 	&		&		&		&		&	1.45	&	0.14	\\
VVVX CL047	&	251.500630	&	-41.766043	&	-3.8320	&	1.9812	&	-4.7850	&	2.1367	&	0.2699	&	0.0540	&	3038	&	2529	&	4043	&	-111.89	&	0.51	&	1.23	&	0.28	\\
VVVX CL049	&	251.860180	&	-41.244139	&	-3.7890	&	1.4063	&	-1.7590	&	2.4079	&	1.1548	&	1.0419	&	765	&	471	&	2173	&		&		&	0.93	&	0.10	\\
VVVX CL050	&	252.328640	&	-48.592074	&	-3.3730	&	0.9842	&	-4.3005	&	2.4106	&	0.3585	&	0.4223	&	1309	&	894	&	2680	&		&		&	1.25	&	0.14	\\
VVVX CL051	&	253.042590	&	-40.065037	&	-3.9830	&	1.8397	&	-3.9330	&	2.1836	&		&		&	 	&		&		&	-87.12	&	24.18	&	1.23	&	0.31	\\
\hline
\end{tabular}
\label{parallaxes}
\end{table}
\end{landscape}

\begin{landscape}\tabcolsep=5pt
\begin{table}
\contcaption{VVVX Cluster Candidates: Parameters}
\begin{tabular}{llcccccccccccccc} % 16 columns, alignment for each
		\hline
\multicolumn{1}{c}{Name}	&	\multicolumn{1}{c}{$\alpha(2000)$}	&	\multicolumn{1}{c}{$\delta(2000)$}	& \multicolumn{1}{c}{$\mu_\alpha\cos\delta$}	&	\multicolumn{1}{c}{Err $\mu_\alpha\cos\delta$}	&	\multicolumn{1}{c}{$\mu_\delta$}	&	\multicolumn{1}{c}{Err $\mu_\delta$}	&	\multicolumn{1}{c}{$\pi$}	&	\multicolumn{1}{c}{Err $\pi$}	&	\multicolumn{1}{c}{Dis} 	&	\multicolumn{1}{c}{Dis 5\%}	&	\multicolumn{1}{c}{Dis 95\%}	&	\multicolumn{1}{c}{RV} 	&	\multicolumn{1}{c}{Err RV}	&	\multicolumn{1}{c}{E(BP-RP)}	&	\multicolumn{1}{c}{Err E(BP-RP)}	\\
\multicolumn{1}{c}{}	&	\multicolumn{1}{c}{$^\circ$}	&	\multicolumn{1}{c}{$^\circ$}	& \multicolumn{1}{c}{mas}	&	\multicolumn{1}{c}{mas}	&	\multicolumn{1}{c}{mas}	&	\multicolumn{1}{c}{mas}	&	\multicolumn{1}{c}{mas}	&	\multicolumn{1}{c}{mas}	&	\multicolumn{1}{c}{kpc} 	&	\multicolumn{1}{c}{kpc}	&	\multicolumn{1}{c}{kpc}	&	\multicolumn{1}{c}{km s$^{-1}$} 	&	\multicolumn{1}{c}{km s$^{-1}$}	&	\multicolumn{1}{c}{mag}	&	\multicolumn{1}{c}{mag}	\\
\hline
VVVX CL052	&	253.294640	&	-39.836432	&	-2.5690	&	2.2240	&	-4.7870	&	1.7466	&	0.4713	&	0.0485	&	2042	&	1791	&	2468	&		&		&	1.44	&	0.03	\\
VVVX CL053	&	255.785070	&	-37.520748	&	-3.5670	&	1.5156	&	-4.1120	&	2.0580	&		&		&	 	&		&		&	-118.28	&	4.16	&	1.32	&	0.14	\\
VVVX CL055	&	259.483430	&	-42.657861	&	-1.9610	&	2.2399	&	-3.0390	&	2.1161	&		&		&	 	&		&		&		&		&	1.41	&	0.10	\\
VVVX CL056	&	259.499020	&	-42.094804	&	-1.0270	&	1.4837	&	-3.1790	&	2.0557	&	0.2695	&	0.0773	&	2740	&	2209	&	3891	&	-21.96	&	6.59	&	1.30	&	0.19	\\
VVVX CL057	&	259.780310	&	-41.405093	&	-2.6640	&	1.5039	&	-5.0680	&	1.8897	&	0.7533	&	0.0419	&	1320	&	1217	&	2811	&		&		&		&		\\
VVVX CL058	&	262.414960	&	-39.466213	&	-2.1975	&	1.7054	&	-4.7495	&	1.9031	&	0.4497	&	0.1134	&	1882	&	1503	&	2810	&		&		&	1.33	&	0.39	\\
VVVX CL059	&	262.844920	&	-38.469595	&	-3.3055	&	2.2459	&	-6.0655	&	1.4374	&		&		&	 	&		&		&		&		&	1.22	&	0.29	\\
VVVX CL060	&	268.852770	&	-18.000075	&	-2.3860	&	2.0186	&	-4.9815	&	1.5989	&	0.2806	&	0.1362	&	2198	&	1684	&	3471	&		&		&	1.08	&	0.17	\\
VVVX CL064	&	270.249460	&	-17.749453	&	-2.4080	&	2.8298	&	-5.1290	&	3.4491	&		&		&	 	&		&		&		&		&	1.15	&	0.35	\\
VVVX CL065	&	270.369110	&	-17.946281	&	-3.0780	&	1.5112	&	-6.1625	&	1.4636	&		&		&	 	&		&		&		&		&	1.32	&	0.27	\\
VVVX CL066	&	271.856700	&	-20.028048	&	-2.7450	&	0.3543	&	-7.7680	&	2.9813	&		&		&	 	&		&		&		&		&		&		\\
VVVX CL076	&	273.354540	&	-18.092123	&	-0.2315	&	1.1797	&	-1.1350	&	1.0050	&	0.7011	&	0.3490	&	1141	&	819	&	2322	&		&		&	1.43	&	0.26	\\
VVVX CL077	&	273.396040	&	-18.096607	&	-1.1500	&	0.5148	&	-0.8470	&	0.4311	&		&		&	 	&		&		&		&		&	1.36	&	0.10	\\
VVVX CL080	&	273.527110	&	-17.934716	&	-0.1565	&	1.4374	&	-3.1130	&	3.6308	&	0.5336	&	0.4174	&	1201	&	829	&	2514	&		&		&	0.97	&	0.21	\\
VVVX CL086	&	274.040020	&	-13.755454	&	-2.3780	&	2.5917	&	-3.1320	&	4.0305	&	0.5628	&	0.0620	&	1718	&	1494	&	2116	&		&		&	1.35	&	0.10	\\
VVVX CL089	&	274.247920	&	-16.999377	&	0.5210	&	1.8414	&	-1.7870	&	3.3595	&	0.6742	&	0.2940	&	1210	&	886	&	2335	&		&		&	1.05	&	0.53	\\
VVVX CL102	&	274.916667	&	-13.362500	&		&		&		&		&		&		&	 	&		&		&		&		&	1.47	&	0.10	\\
VVVX CL103	&	274.923010	&	-21.059604	&	-3.1400	&	3.2674	&	-1.0220	&	3.3342	&	0.8086	&	0.1970	&	1154	&	907	&	1894	&		&		&		&		\\
VVVX CL107	&	275.250670	&	-18.696785	&	-2.0890	&	2.2319	&	-4.0075	&	1.9944	&		&		&	 	&		&		&		&		&	1.09	&	0.13	\\
VVVX CL111	&	276.365960	&	-12.853166	&	-3.0080	&	1.4469	&	-5.4595	&	2.9577	&		&		&	 	&		&		&		&		&		&		\\
VVVX CL112	&	276.389290	&	-11.588813	&	-1.7930	&	1.4948	&	-3.0460	&	2.3819	&	0.3824	&	0.2343	&	1621	&	1186	&	2895	&		&		&	1.45	&	0.31	\\
\hline
\end{tabular}
%\label{parallaxes}
\end{table}
\end{landscape}

%\begin{landscape}\tabcolsep=5pt
%\begin{table}
%\contcaption{VVVX Cluster Candidates: Parameters}
%\begin{tabular}{llcccccccccccccc} % 16 columns, alignment for each
%		\hline
%\multicolumn{1}{c}{Name}	&	\multicolumn{1}{c}{$\alpha(2000)$}	&	\multicolumn{1}{c}{$\delta(2000)$}	& \multicolumn{1}{c}{$\mu_\alpha\cos\delta$}	&	\multicolumn{1}{c}{Err $\mu_\alpha\cos\delta$}	&	\multicolumn{1}{c}{$\mu_\delta$}	&	\multicolumn{1}{c}{Err $\mu_\delta$}	&	\multicolumn{1}{c}{$\pi$}	&	\multicolumn{1}{c}{Err $\pi$}	&	\multicolumn{1}{c}{Dis} 	&	\multicolumn{1}{c}{Dis 5\%}	&	\multicolumn{1}{c}{Dis 95\%}	&	\multicolumn{1}{c}{RV} 	&	\multicolumn{1}{c}{Err RV}	&	\multicolumn{1}{c}{E(BP-RP)}	&	\multicolumn{1}{c}{Err E(BP-RP)}	\\
%\multicolumn{1}{c}{}	&	\multicolumn{1}{c}{$^\circ$}	&	\multicolumn{1}{c}{$^\circ$}	& \multicolumn{1}{c}{mas}	&	\multicolumn{1}{c}{mas}	&	\multicolumn{1}{c}{mas}	&	\multicolumn{1}{c}{mas}	&	\multicolumn{1}{c}{mas}	&	\multicolumn{1}{c}{mas}	&	\multicolumn{1}{c}{kpc} 	&	\multicolumn{1}{c}{kpc}	&	\multicolumn{1}{c}{kpc}	&	\multicolumn{1}{c}{km s$^{-1}$} 	&	\multicolumn{1}{c}{km s$^{-1}$}	&	\multicolumn{1}{c}{mag}	&	\multicolumn{1}{c}{mag}	\\
%\hline
%\hline
%\end{tabular}
%\label{parallaxes}
%\end{table}
%\end{landscape}

\newpage

\section{Composite color images of VVVX open cluster candidates}

Figure\,\ref{rgb_all} shows the composite color images of VVVX open cluster candidates. The field of view is 2.5\,$\times$\,2.5\,arcmin, north is up, east to the left (unless specified in the image). The clusters are shown in order corresponding of the Table A1.  
The red large circles indicate the cluster candidate approximated boundaries and are given to assist the reader to easy identification.

\begin{figure*}
\begin{center}
\includegraphics[width=14.0cm]{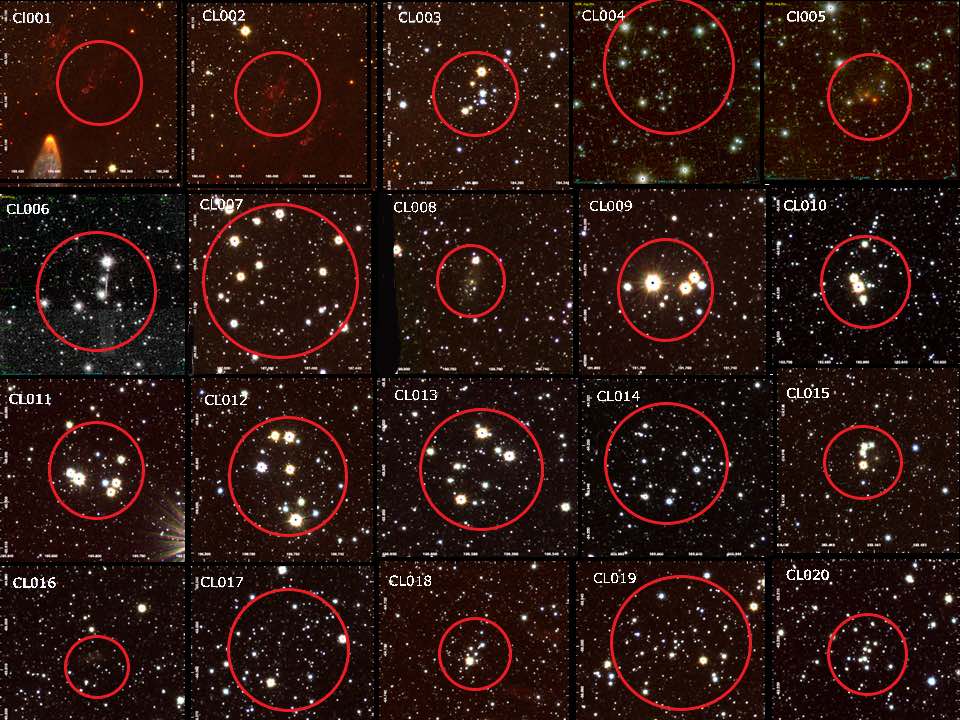}
\includegraphics[width=14.0cm]{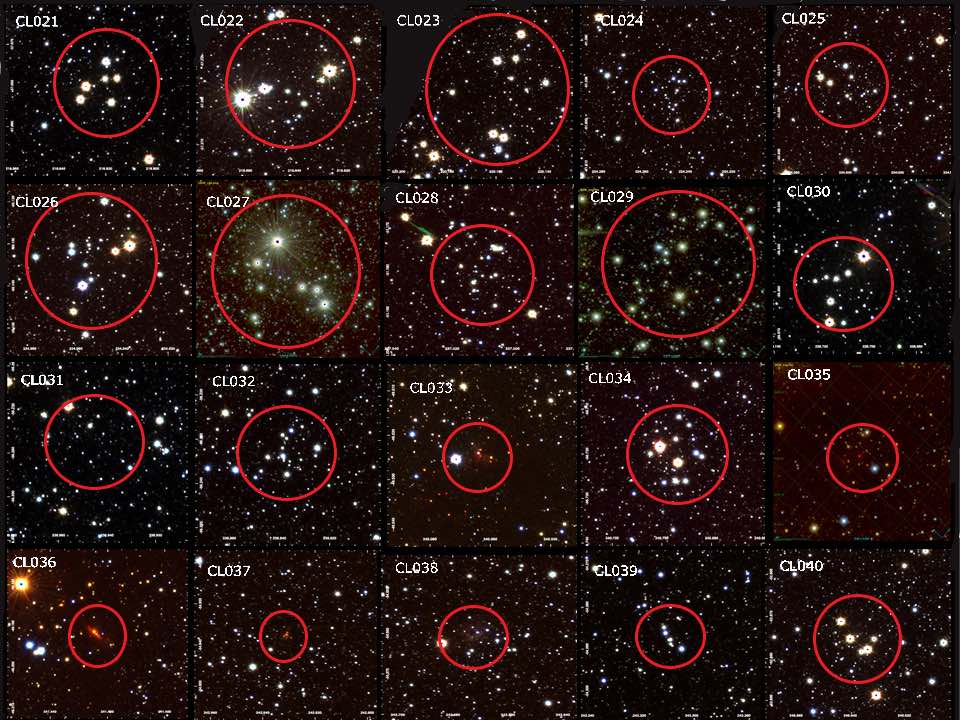}
\end{center}
\caption{VVVX $JHK_S$ composite color images of the open cluster candidates. The field of view is 2.5\,$\times$\,2.5\,arcmin, north is up, east to the left.}
\label{rgb_all}
\end{figure*}

\begin{figure*}
\begin{center}
\includegraphics[width=14.0cm]{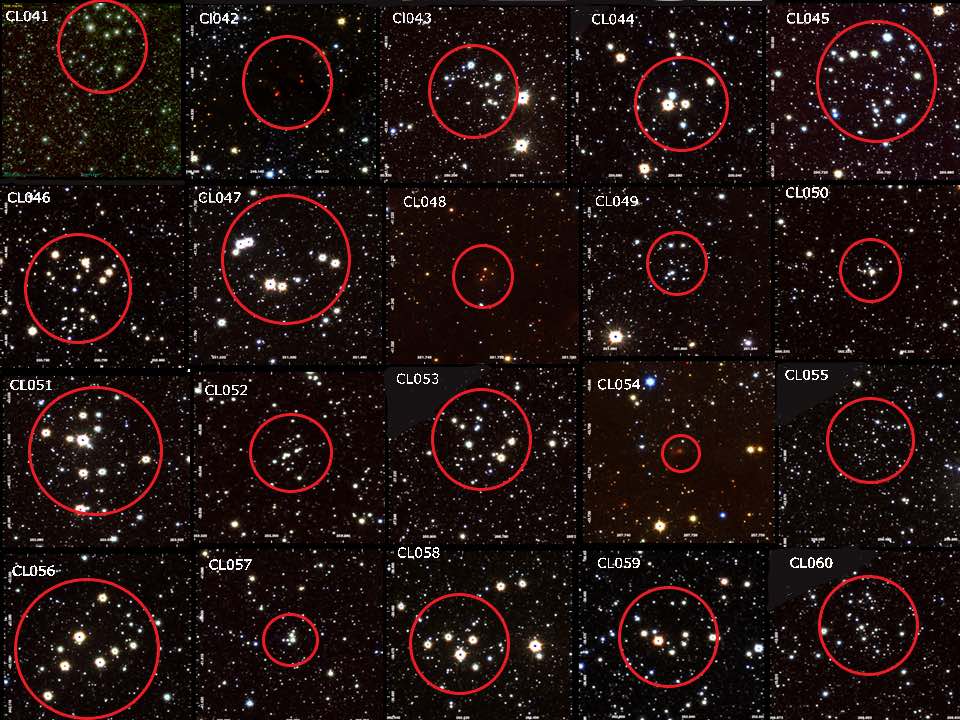}
\includegraphics[width=14.0cm]{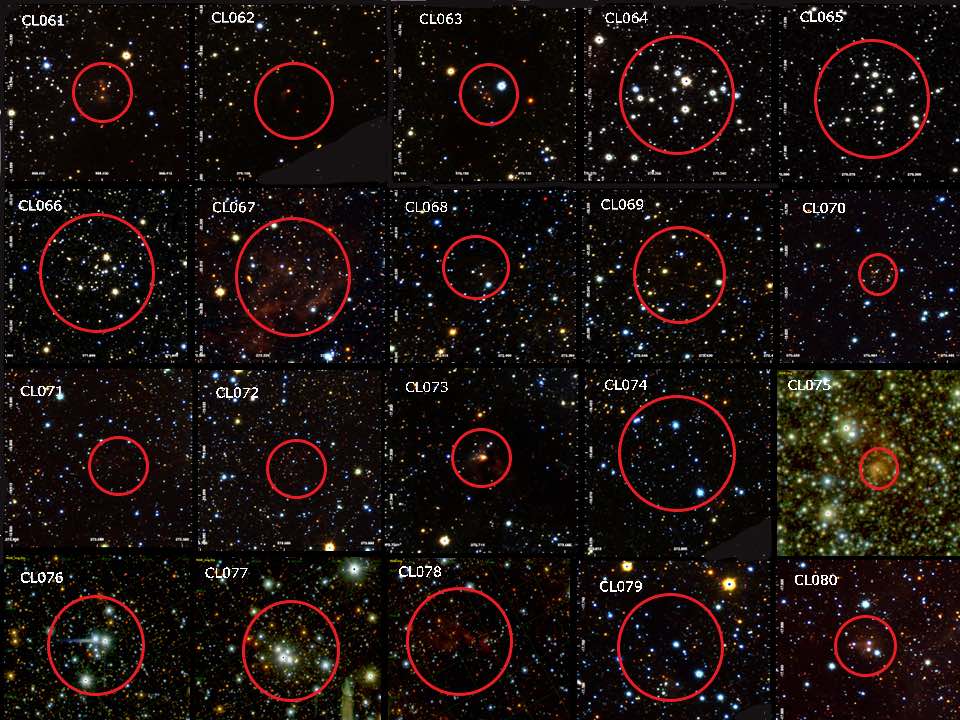}
\end{center}
\contcaption{VVVX $JHK_S$ composite color images of the open cluster candidates. }
%\label{rgb_all1}
\end{figure*}

\begin{figure*}
\begin{center}
\includegraphics[width=14.0cm]{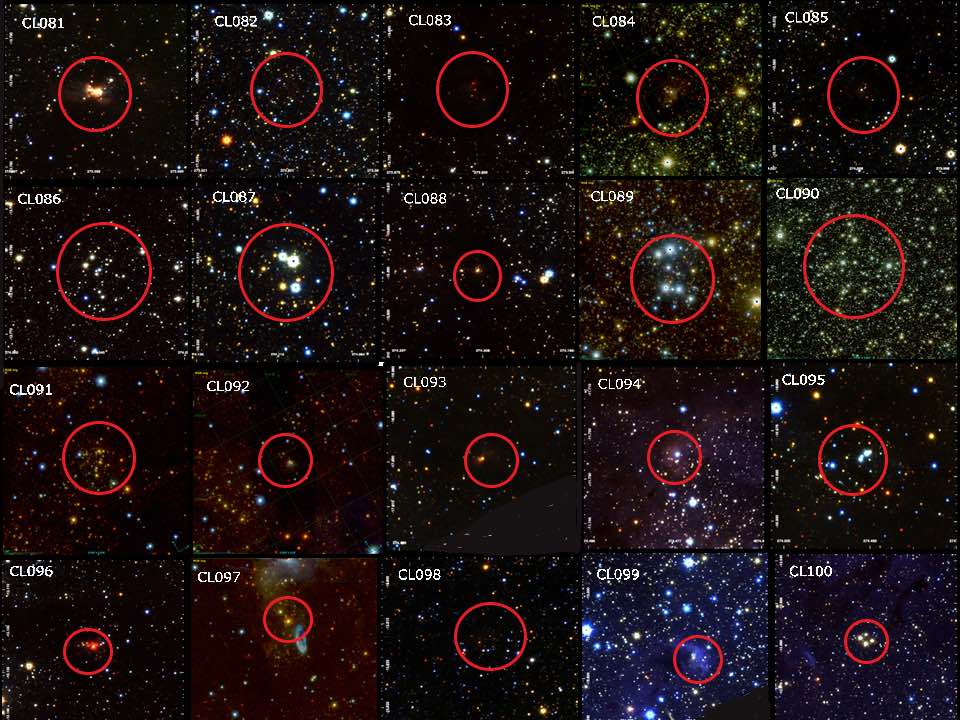}
\includegraphics[width=14.0cm]{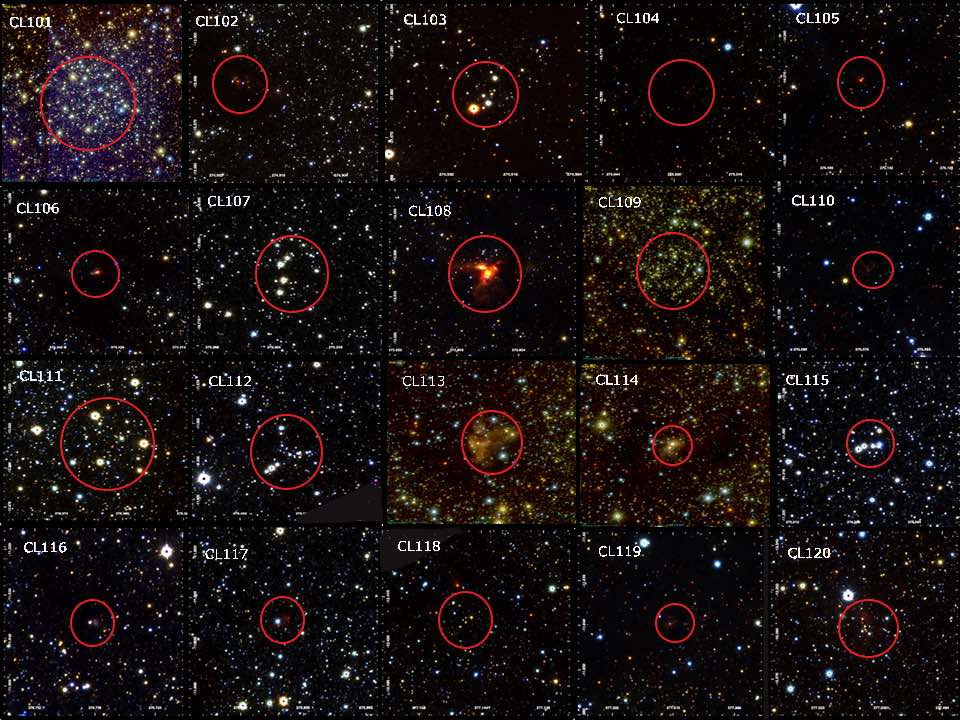}
\end{center}
\contcaption{VVVX $JHK_S$ composite color images of the open cluster candidates. }
%\label{rgb_all2}
\end{figure*}

\newpage

\section{Color magnitude diagrams of some of the VVVX open cluster candidates.}

The color magnitude diagrams of some of the new open cluster candidates are plotted in Fig.\,\ref{all_cmd}.

\begin{figure*}
\includegraphics[width=17.6cm]{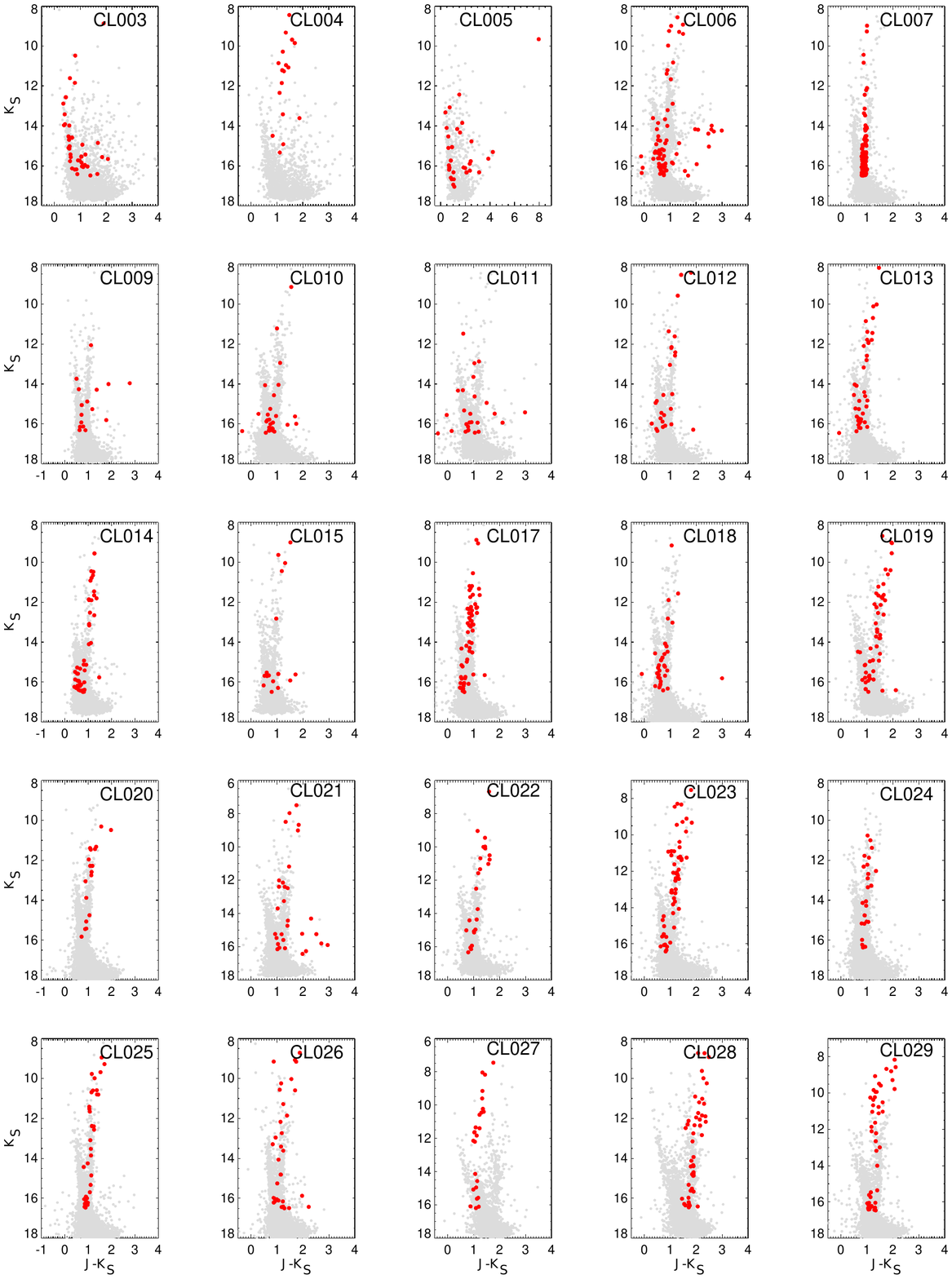}
\caption{VVVX $K_{S}$ vs $(J-K_{S})$ color-magnitude diagrams of the new open cluster candidates. The stars within 2.5 armin radius are plotted with grey points, the large red circles stand for probable cluster members. }
\label{all_cmd}
\end{figure*}

\begin{figure*}
\includegraphics[width=17.6cm]{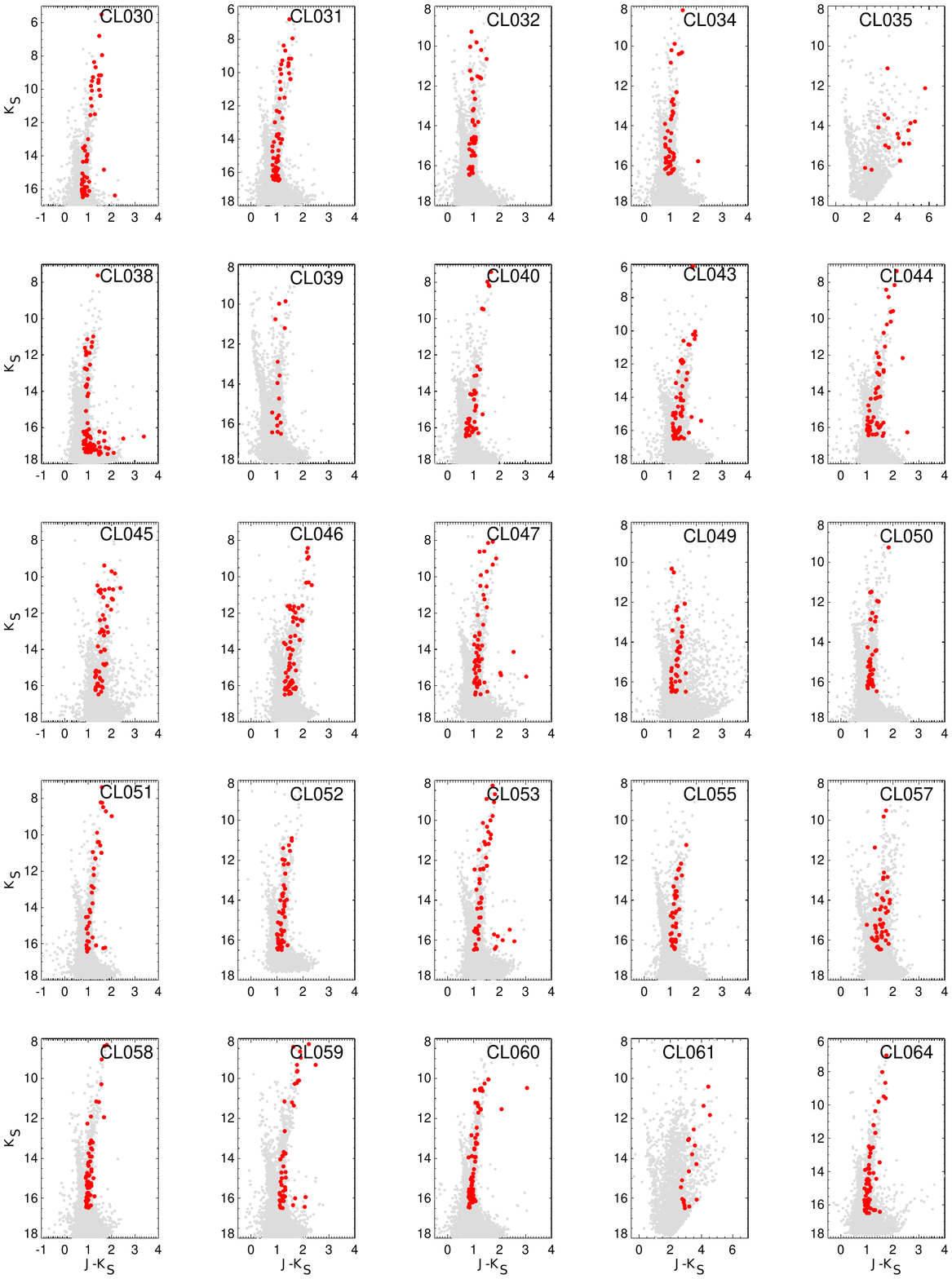}
\contcaption{VVVX $K_{S}$ vs $(J-K_{S})$ color-magnitude diagrams of the new open cluster candidates. }
%\label{all_cmd}
\end{figure*}

\begin{figure*}
\includegraphics[width=17.6cm]{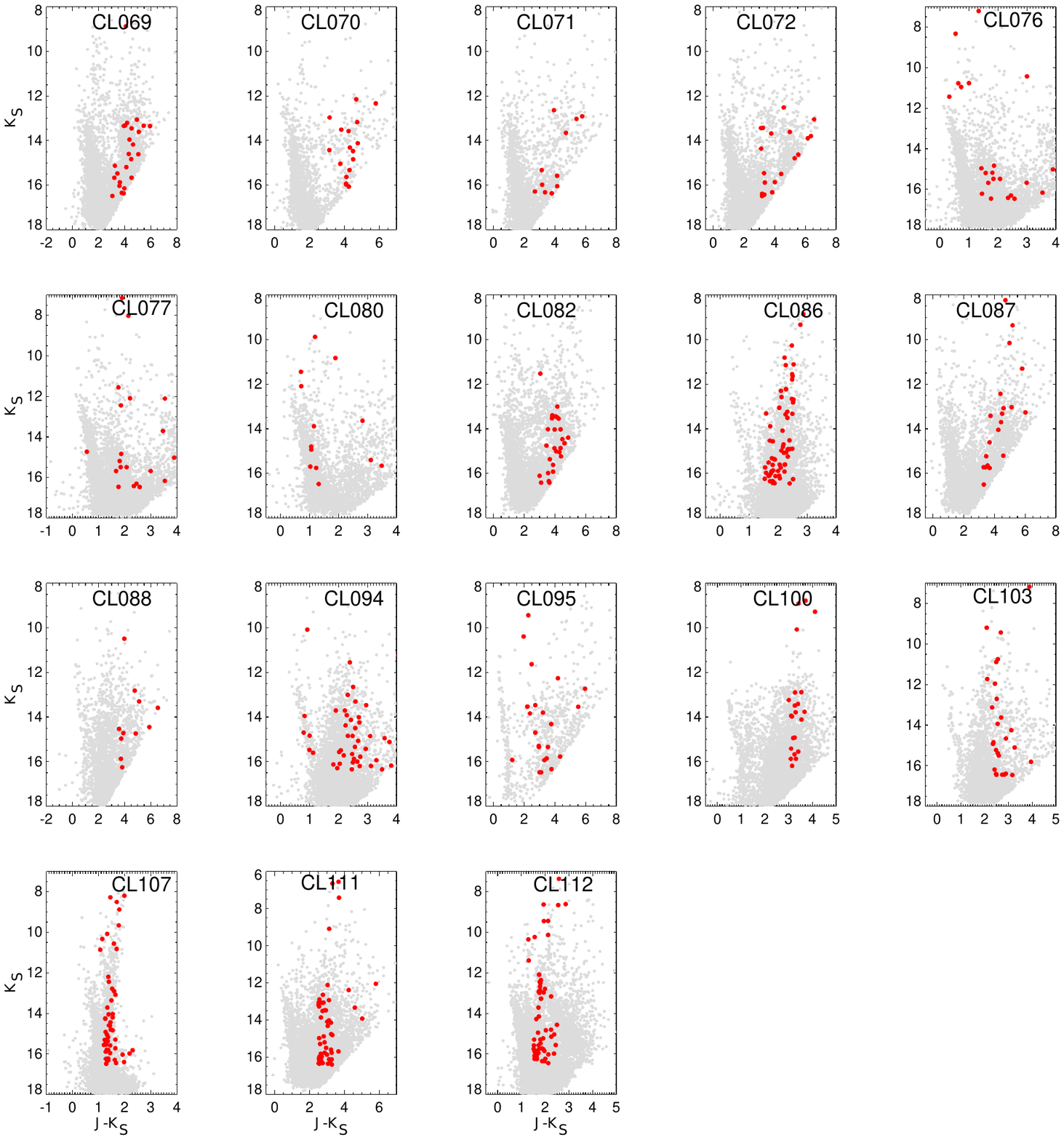}
\contcaption{VVVX $K_{S}$ vs $(J-K_{S})$ color-magnitude diagrams of the new open cluster candidates.}
%\label{all_cmd}
\end{figure*}


%%%%%%%%%%%%%%%%%%%%%%%%%%%%%%%%%%%%%%%%%%%%%%%%%%

% Don't change these lines
\bsp	% typesetting comment
\label{lastpage}
\end{document}